\documentclass[lettersize,journal]{IEEEtran}
\usepackage[table]{xcolor}
\usepackage{balance}
\usepackage{booktabs}
\usepackage{graphicx}
\usepackage[caption=false,font=normalsize,labelfont=sf,textfont=sf]{subfig}
\usepackage{blindtext}
\usepackage{adjustbox}
\usepackage{multirow}
\usepackage{tabularx}
\usepackage{tikz}
\usepackage{listings}
\usepackage{etoolbox}
\usepackage{url}
\usepackage{setspace}
\usepackage[british,english]{babel}
\usepackage{enumitem}
\usepackage{amsmath,amsfonts,amssymb}
\usepackage{array}
\usepackage{tcolorbox}
\usepackage{textcomp}
\usepackage{soul}
\usepackage{diagbox}
\usepackage{bbding}
\usepackage{bibentry}
\usepackage{ifthen}
\usepackage{xspace}
\usepackage{algorithmic}
\usepackage{stfloats}
\usepackage{verbatim}

\definecolor{bestgreen}{HTML}{DDE9B2}
\definecolor{secondblue}{HTML}{DCE6F1}
\newcommand{\best}[1]{\cellcolor{bestgreen}\textbf{#1}}
\newcommand{\second}[1]{\cellcolor{secondblue}\textbf{#1}}

\definecolor{Gray}{gray}{0.9}

\def\BibTeX{{\rm B\kern-.05em{\sc i\kern-.025em b}\kern-.08em
    T\kern-.1667em\lower.7ex\hbox{E}\kern-.125emX}}

\newcommand{\SystemName}{\textsc{ATLAS}\xspace}
\newcommand\cparagraph[1]{\vspace{1.5mm} \noindent \textbf{#1}\xspace}

\newcommand{\circled}[2][]{\tikz[baseline=(char.base)]
	{\node[shape = circle, draw, inner sep = 1pt]
		(char) {\phantom{\ifblank{#1}{#2}{#1}}};\node at (char.center) {\makebox[0pt][c]{#2}};}}
\robustify{\circled}

\begin{document}

\title{\SystemName: A Multi-Agent Framework for Lifting Binary to Canonical Compiler IR}

	\author{

    Xiaoao Zhu, Ni Zhang,
Jie Ren*,~\IEEEmembership{Member,~IEEE}, Zhiqiang Li, Jie Zheng,
   Zhanyong Tang~\IEEEmembership{Member,~IEEE},\\
   Zheng Wang~\IEEEmembership{Member,~IEEE}\\
\thanks{
This work was supported in part by the National Natural Science Foundation of China under Grant 62372281 and in part by the Fundamental Research Funds for the Central Universities of China under Grant GK202501010 (Corresponding author: Jie Ren.)
    
Xiaoao Zhu, Ni Zhang,
Jie Ren and Zhiqiang Li are with the School
of Artificial Intelligence and Computer Science, Shaanxi Normal University. Xi’an, Shaanxi, China. (Xiaoao Zhu and Ni Zhang are co-first authors. ) 

Jie Zheng, Zhanyong Tang, with the School of Computer Science, Northwest University, Xian, China. 

Zheng Wang is with the School of Computing, University of Leeds, LS2 9JT Leeds, U.K.

E-mail: {renjie@snnu.edu.cn}.
}}

\markboth{Journal of \LaTeX\ Class Files,~Vol.~18, No.~9, September~2020}{How to Use the IEEEtran \LaTeX \ Templates}

\maketitle
	
	\begin{abstract}

Lifting stripped, heavily optimized binaries to a canonical, pre-optimization LLVM intermediate representation (IR) is essential for downstream security analysis and legacy software migration when source code is unavailable.  However, aggressive compiler optimizations rewrite control and data flow, remove structural metadata, and introduce type-erased low-level idioms, causing current lifters to produce malformed or semantically divergent IR.

 We present \SystemName, a multi-agent framework designed to reconstruct formally valid and canonical LLVM IR from optimized disassembly. \SystemName combines (i) Structure-aware, region-level retrieval, which utilizes pseudo-probe attribution to align optimized instructions with canonical IR exemplars, and (ii) a collaborative refinement loop that closes the gap between probabilistic generation and deterministic compiler constraints. Concretely, a \textsc{Lift} agent generates an initial IR draft conditioned on retrieved exemplars, an \textsc{Advisor} agent interprets LLVM verifier/compilation diagnostics and static checks to propose localized repair actions, and a \textsc{Validation} agent applies the edits, re-runs toolchain verification and (when available) runtime checks, and feeds the resulting diagnostics back to the \textsc{Advisor}. The agents iterate until the IR is verifier-clean and behaviorally consistent, or a budget limit is reached.

 On HumanEval-Decompile and MBPP, \SystemName outperforms seven baselines on lifting LLVM IR, achieving 91.8\% re-executability and 89.4\% CFG similarity across ARM, x86-64, and RISC-V binaries.  Moreover, our evaluation demonstrates that \SystemName also outperforms the Binary to C to IR pipelines in pointer-flow precision and downstream analyzability. By recovering a well-formed, machine-centric representation, \SystemName enables practical cross-ISA recompilation and verifiable binary-only migration workflows.
\end{abstract}


\section{Introduction}

Recovering usable program representations from binaries is important for software security analysis, legacy-system maintenance, and migration across heterogeneous computing platforms. In these settings, source code may be unavailable, while the deployed executable remains available for inspection and testing. Binary recovery must therefore provide a representation that supports both reasoning about program behavior and subsequent transformation. Although much recent decompilation research targets human-readable source code~\cite{tan2024llm4decompiledecompilingbinarycode,tan2025sk2decompile}, downstream compiler workflows also require explicit control flow, consistent types, and well-defined memory operations.

\begin{figure}[t!]
    \centering
    \includegraphics[width=0.5\textwidth]{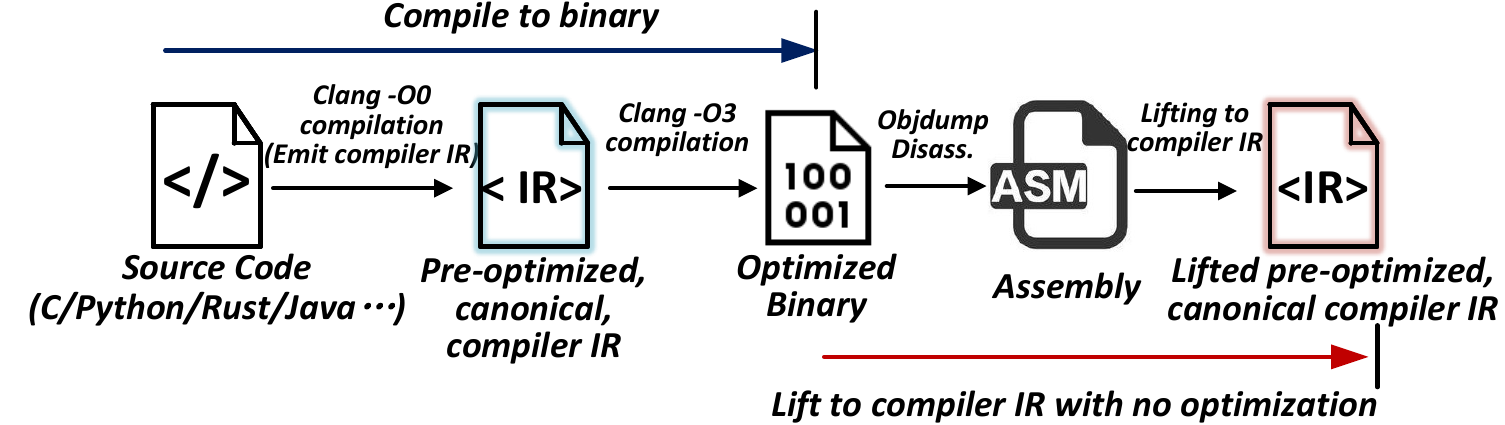}
    \vspace{-4mm}
    \caption{The compilation pipeline from source code to optimized binaries and the reverse lifting task, which targets canonical, pre-optimization compiler IR.}
    \vspace{-5mm}
    \label{fig:compflow}
\end{figure}

Compiler intermediate representation (IR) provides a useful target for these workflows. LLVM IR can be checked by the LLVM verifier, consumed by program-analysis tools, and recompiled using different compiler backends. Figure~\ref{fig:compflow} illustrates the lifting task considered in this paper: reconstructing canonical, pre-optimization LLVM IR from optimized binaries. We use \emph{canonical} as an operational reference, defined by the unoptimized IR emitted under a fixed \texttt{clang -O0} toolchain configuration. This reference provides a consistent target for structural recovery; it does not imply that an optimized binary uniquely determines its original IR. The practical objective is to recover well-formed IR that preserves the binary's behavior on the evaluated inputs and remains useful for analysis and recompilation.

This recovery task is difficult because compiler optimizations alter control and data flow, eliminate variables, and replace structured computations with architecture-specific instruction sequences. Rule-based lifters, including LLVM-MCTOLL~\cite{10.1145/3316482.3326354} and RetDec~\cite{retdec}, translate machine instructions and reconstruct program structure through deterministic rules and analysis. Their recovered representations can nevertheless differ substantially from compiler-generated IR. Large language models (LLMs) provide a complementary means of inferring code structure, but plausible output can still contain invalid types, inconsistent static single assignment (SSA) definitions, or incorrect computations. Successful lifting consequently requires both structural checks and behavioral evidence.

Our prior work, BRIDGE~\cite{zhu2026bridge}, combines control-flow-aware retrieval with iterative verification for binary-to-canonical-IR lifting. It uses pseudo-probe attribution to construct aligned assembly--IR exemplars and incorporates static diagnostics and runtime feedback into refinement. \SystemName builds on this foundation and develops the repair and validation workflow further. Its focus is on localizing repairs, obtaining behavioral evidence when developer-written tests are unavailable, and assessing whether the resulting IR supports reuse across instruction set architectures (ISAs).

Three challenges motivate these extensions. \textbf{First, compiler diagnostics must be translated into appropriately scoped repairs.} A reported type error or malformed \texttt{phi} instruction can involve definitions, uses, and predecessor blocks beyond the reported line. Repair therefore requires identifying the relevant dependencies and enforcing their constraints together. Regenerating a function from an error log provides little control over which valid parts of the current candidate are changed. A more targeted workflow must connect each diagnostic to an affected IR region and a concrete repair action.

\textbf{Second, behavioral feedback requires an executable interface and a usable reference.} Curated benchmarks supply function signatures and unit tests, whereas binary-only settings may provide neither. Generating test inputs alone is insufficient: the system must infer how to call the function, construct a feasible harness, and obtain reference observations from the original executable. Incorrect interface assumptions can make test failures uninformative. The resulting observations must also be interpreted within their coverage, since passing a finite set of tests does not establish equivalence for all inputs.

\textbf{Third, successful execution on the source architecture does not establish cross-ISA reusability.} Recovered IR may retain assumptions about data layout, external interfaces, or machine-specific operations. Recompiling and testing it on another architecture therefore provides an additional assessment of its utility. This assessment complements structural similarity and same-architecture execution by examining whether recovered computations can be reused through a different compiler backend.

We present \SystemName, a multi-agent framework that organizes lifting around three explicit responsibilities: initial synthesis, repair planning, and patch validation. The \textsc{Lift Agent} uses CFG-aligned retrieval to generate an initial IR candidate. The \textsc{Advisor Agent} interprets compiler, verifier, and runtime diagnostics, identifies the affected instructions and dependencies, and proposes localized repair actions. Its repair context includes predecessor constraints for \texttt{phi} nodes, def--use slices for inconsistent values, and available binary-interface information for external calls. The \textsc{Validation Agent} applies these actions and re-runs toolchain checks and available behavioral tests. Failures are returned with the updated candidate for further diagnosis, with refinement bounded by a maximum of five repair iterations. This organization makes the diagnostic evidence, proposed changes, and validation outcomes explicit at each step.

To support refinement without developer-written tests, \SystemName also constructs behavioral checks from the original binary. It combines static interface inference with observations from dynamic instrumentation, uses these inputs to propose candidate function signatures and test harnesses, and screens candidates through consistency checks and fuzzing-based execution. For accepted harnesses, it compares the original binary and the recompiled IR on the same inputs, including return values and observable memory effects. These differential observations provide concrete evidence for subsequent repairs without requiring source-level expected outputs.

We evaluate \SystemName on HumanEval-Decompile~\cite{tan2024llm4decompiledecompilingbinarycode} and MBPP~\cite{austin2021program}, covering x86-64, ARM64, and RISC-V binaries compiled at optimization levels \texttt{-O0} through \texttt{-O3}. The evaluation includes rule-based and general-purpose LLM baselines where supported, studies retrieval granularity, refinement, and test availability, and examines downstream SVF-based analysis~\cite{sui2016svf} and cross-ISA migration. On HumanEval-Decompile, \SystemName achieves 94.5\% re-executability for x86-64 inputs, averaged across optimization levels. In the x86-64-to-ARM64 migration experiment, it achieves 93.6\% re-executability on the target architecture. These results characterize lifting quality and downstream utility under the evaluated configurations.

This paper makes the following contributions:

\begin{itemize}
    \item We present \SystemName, a multi-agent framework that separates synthesis, repair planning, and validation for lifting optimized, stripped binaries to canonical LLVM IR.
    \item We support structure-aware retrieval across multiple ISAs using aligned assembly--IR exemplars and LoRA adaptation of the assembly encoder for ARM64 and RISC-V.
    \item We develop localized repair that maps diagnostics to affected IR slices and derives type, def--use, and control-flow constraints to guide targeted edits.
    \item We develop a behavioral validation workflow combining interface inference, dynamic observations, harness screening, and differential execution to guide repairs without developer-written tests.
    \item We evaluate \SystemName across three ISAs and four optimization levels, examining lifting quality, component effectiveness, downstream analysis, and cross-ISA migration.
\end{itemize}

\begin{figure*}[t!]
 	\begin{center}
 		\includegraphics[width=1\textwidth]{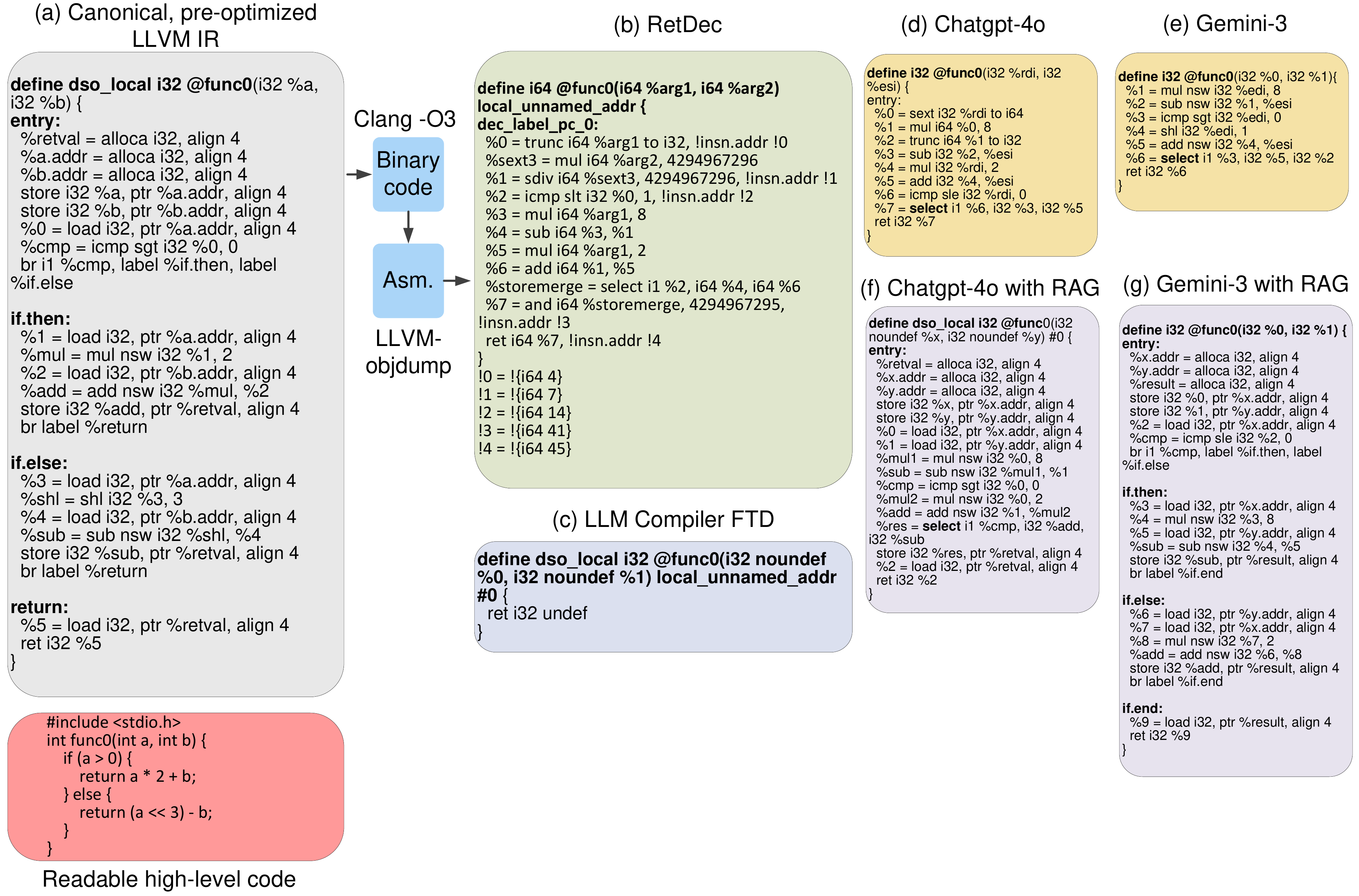}
 	\end{center}
 	\caption{Lifting results of different approaches (b–g) compared with the ground-truth canonical, pre-optimized LLVM IR (a). The rule-based lifter RetDec distorts the original \texttt{if-else} control-flow structure. The general-purpose LLMs, especially Gemini-3 with RAG, better preserve the CFG and produce IR closer to the reference.}
 	 \vspace{-4mm}
 	\label{fig:moti}
 \end{figure*}

\section{Motivation}

To motivate our work, we present a task of lifting an \texttt{-O3} optimized binary back to its corresponding pre-optimized, canonical LLVM IR (\texttt{-O0}). 
\textit{In this paper, canonical is an operational target: we define it as the unoptimized LLVM IR emitted by a fixed clang -O0 toolchain configuration.}
Figure~\ref{fig:moti}a presents the canonical, pre-optimized LLVM IR source code, which uses conditional \texttt{if-else} control flow to perform arithmetic operations, either doubling or bit-shifting variable \texttt{a} and combining the result with variable \texttt{b}.  
Specifically, we first compile the LLVM IR with Clang (version 19)~\cite{llvmproject} using the \textbf{-O3} flag to generate an optimized binary. We then use \textbf{llvm-objdump}~\cite{llvmproject} to obtain the assembly instructions, which serve as input to different lifters for reconstructing the original LLVM IR.

Figure~\ref{fig:moti} compares the lifting results of six approaches for converting \texttt{-O3} optimized assembly code to pre-optimized, canonical LLVM IR (\texttt{-O0}). These include the LLVM-based decompiler RetDec~\cite{retdec}, the Meta fine-tuned LLM Compiler FTD~\cite{cummins2024metalargelanguagemodel}, and two typical general-purpose LLMs: ChatGPT-4o~\cite{chatgpt2025} and Gemini-3~\cite{GoogleGemini}. Additionally, we explore two variants that integrate RAG with two general-purpose LLMs. RAG retrieves similar code segments from the ExeBench dataset~\cite{10.1145/3520312.3534867}  to construct contextual prompts. For the two general-purpose LLMs, we use the prompt: \textit{\{Convert the following llvm-objdump assembly (compiled with -O3) into canonical LLVM IR that matches the unoptimized clang -O0 IR style and passes llvm-as/opt -verify. Do not apply optimizations\}}. Table~\ref{tab:decomp_comparison} reports the lifting performance using four metrics:
\begin{itemize}

    \item \textbf{Re-executable} evaluates whether the reconstructed LLVM IR executes all test cases correctly, ensuring functional equivalence to the original program.
    \item \textbf{Edit Similarity}~\cite{armengolestape2024sladeportablesmalllanguage} measures the similarity between the decompiled LLVM IR and the ground-truth LLVM IR. It yields a score between 0 and 1, where higher values indicate better syntactic and semantic preservation.
    \item \textbf{CFG Match} evaluates whether the reconstructed LLVM IR recovers the original Control Flow Graph from the assembly's control flow.
    \item \textbf{PCI (Pointer Coverage Index)} measures how much pointer-related structure in the lifted IR can be recovered by SVF. Higher is better, indicating clearer pointer types/flows and more complete downstream analysis.
\end{itemize}

 \begin{table}[t!]
    \centering
    \scriptsize
            \caption{Comparison of different decompilation methods in edit similarity, test case pass, and branch recovery on x86-64.}
    \renewcommand{\arraystretch}{1.05}
    \setlength{\tabcolsep}{3pt} \rowcolors{2}{gray!10}{white}

    \vspace{-2mm}
    \begin{tabular}{lrrrr}
        \toprule
        \textbf{Method} &\textbf{Re-executable} 
&\textbf{Edit Sim. (\%)}  
& \textbf{CFG Match.}  &\textbf{PCI}\\
        \midrule
        RetDec               &\Checkmark   &38.2  
& \XSolidBrush  &25\\
        LLM Compiler FTD     &\XSolidBrush &11.0  
& \XSolidBrush  &0\\
        ChatGPT-4o           &\Checkmark   &32.0  
& \XSolidBrush  &18\\
        ChatGPT-4o+RAG       &\Checkmark   &65.0  
& \XSolidBrush  &32\\
        Gemini-3&\Checkmark   &26.0  
& \XSolidBrush  &21\\
        \textbf{Gemini3+RAG}&\Checkmark &\textbf{82.5}  & \Checkmark  &\textbf{39}\\
        \bottomrule
    \end{tabular}

    \label{tab:decomp_comparison}
    \vspace{-3mm}
\end{table}

\cparagraph{RetDec}~\cite{retdec} produces compilable and test-passing IR, but the result can remain far from the canonical pre-optimization LLVM IR targeted in this work. In Figure~\ref{fig:moti}, RetDec passes the available tests, yet achieves only 38.2\% edit similarity, does not recover the original CFG exactly, and attains a low Pointer Coverage Index (PCI = 25). Concretely, the lifted IR replaces the original branch-based structure with a \texttt{select}-based merged computation and introduces low-level arithmetic normalization and masking, which obscures the original control-flow and data-flow organization. As a result, although the output is executable, it is structurally misaligned with the reference IR and substantially less suitable for downstream static analysis.

\cparagraph{LLM Compiler FTD}~\cite{cummins2024metalargelanguagemodel} is a fine-tuned model for reconstructing LLVM IR from assembly text. However, under the raw disassembly setting studied in this paper, its outputs can exhibit substantial behavioral errors. In our motivating example, the model introduces \texttt{undef} and fails to recover key arithmetic operations and the conditional (\texttt{if-else}) structure implied by the assembly, leading to incorrect execution.
A primary reason is \textit{a mismatch in the input domain.} LLM Compiler FTD, as well as related approaches such as Forklift~\cite{armengol-estape2024forklift}, is trained on compiler-emitted assembly (e.g., \texttt{clang -S}), which retains structured labels, symbol names, and other high-level cues. In contrast, raw-binary disassembly from \texttt{llvm-objdump} is typically stripped of symbols and exposes lower-level artifacts, including indirect control flow and ABI-specific details, which makes faithful lifting to canonical \texttt{-O0} IR significantly more challenging.

 \begin{figure*}[t!]
 	\begin{center}
 		\includegraphics[width=0.9\textwidth]{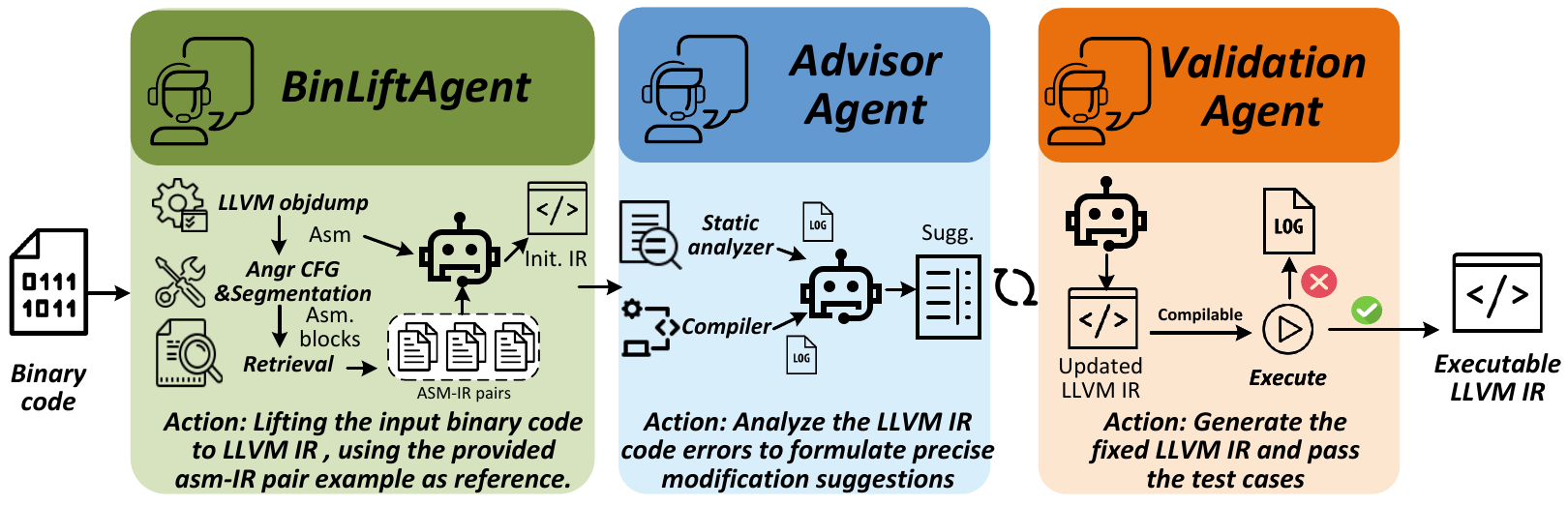}
 	\end{center}
  	\vspace{-3mm}
 	\caption{Overview of \SystemName}
 	 \vspace{-6mm}
 	\label{fig:overview}
 \end{figure*}

\cparagraph{General-Purpose LLMs.}
We observe that both ChatGPT-4o (Figure~\ref{fig:moti}d) and Gemini-3 (Figure~\ref{fig:moti}e) have limitations when lifting raw disassembly to canonical, pre-optimized LLVM IR. First, even when the underlying arithmetic is correct and the outputs pass the available test cases, the generated IR often diverges from the target pre-optimization form, with edit similarity scores of 32.0\% and 26.0\%, respectively. In particular, ChatGPT-4o replaces the original branch-based \texttt{if-else} structure with a \texttt{select}-based formulation, collapsing the CFG into a single block. Although this if-converted IR remains executable, it no longer matches the canonical \texttt{-O0} control-flow structure and is therefore less suitable for downstream CFG-sensitive analyses. This loss of structural fidelity is also reflected in weak downstream analyzability, with low PCI values of 18 for ChatGPT-4o and 21 for Gemini-3.

\cparagraph{General-Purpose LLMs with Our RAG Scheme.}
To evaluate the impact of retrieval augmentation on LLVM IR reconstruction, we integrate our control-flow-aware RAG module with both general-purpose LLMs. The retrieved context provides semantically related assembly--IR exemplars, allowing the models to ground generation in canonical compiler patterns rather than relying solely on open-ended sequence generation. As shown in Figure~\ref{fig:moti}(f)--(g), RAG substantially improves both structural fidelity and downstream analyzability. For ChatGPT-4o, RAG raises the edit similarity from 32.0\% to 65.0\% and improves the PCI from 18 to 32. Although the generated IR still uses a \texttt{select}-based formulation rather than fully recovering the original branch-based CFG, the retrieved exemplars guide the model toward a more canonical stack-based \texttt{-O0}-style structure, with explicit local allocations and clearer memory accesses. This results in IR that is closer to the reference and more suitable for downstream analysis. The improvement is even more pronounced for Gemini-3. After integrating RAG, the model produces compilable and re-executable LLVM IR that recovers the original \texttt{if-else} control-flow structure exactly. This increases edit similarity from 26.0\% to 82.5\%, enables exact CFG matching, and raises PCI from 21 to 39. These results show that retrieval not only reduces structural hallucinations, but also provides the missing control-flow and memory-structure cues needed for reconstructing canonical, analysis-friendly IR from optimized binary disassembly.

 \cparagraph{Insight.}Lifting optimized binary code back into canonical, pre-optimized LLVM IR is challenging, primarily due to aggressive compiler optimizations (e.g., \texttt{-O3}) that obscure high-level control structures by replacing them with low-level, performance-oriented instructions. The motivation reveals that binary-to-LLVM-IR lifting cannot be evaluated solely by compilability or test passing. Rule-based lifters and general-purpose LLMs may produce executable outputs, yet the recovered IR can still deviate substantially from the canonical pre-optimization form, fail to preserve the original control-flow organization, and expose weak pointer/value-flow structure to downstream analyzers. This gap is especially visible in PCI and CFG-based metrics, which show that behaviorally plausible IR may still be structurally unsuitable for compiler and security analyses. At the same time, retrieval-augmented generation substantially improves both structural fidelity and analyzability, suggesting that the missing information is not only lexical but also control- and memory-structural. These observations motivate \SystemName: a lifting framework that combines control-flow-aware retrieval with verification-guided refinement to recover LLVM IR that is not only executable, but also canonical and analysis-friendly.

\section{Overview}
Figure~\ref{fig:overview} overviews \SystemName, a multi-agent framework that lifts optimized binaries to LLVM IR in a canonical pre-optimization form (i.e., the unoptimized IR produced by a fixed \texttt{clang -O0} toolchain configuration). The pipeline comprises three agents: \textsc{Lift}, \textsc{Advisor}, and \textsc{Validation}, each with a distinct and non-overlapping responsibility.

Given a binary, the \textsc{Lift Agent} first disassembles it with \texttt{llvm-objdump} and recovers a coarse function-level CFG using \texttt{angr}~\cite{shoshitaishvili2016state}. It applies a CFG-aligned segmentation procedure to partition the disassembly into control-structure regions (Section~\ref{sec:cfg-seg}). Each region is encoded as a retrieval query against a structure-aware RAG database of aligned (\textit{assembly}, \textit{LLVM IR}) exemplars (Section~\ref{sec:rag}). Conditioned on the retrieved exemplars, the agent invokes an LLM to produce an initial LLVM IR draft, which is passed to the refinement loop.

The \textsc{Advisor Agent} checks the draft with the LLVM toolchain (\texttt{llvm-as}, \texttt{opt -verify}) and a dedicated static analyzer implemented as an LLVM pass. It converts diagnostics into concrete repair actions, including type corrections, control-flow consistency fixes, and merge/value-definition repairs.

The \textsc{Validation Agent} applies the proposed edits and re-runs compilation and validation. When test inputs are available, it also executes the recompiled artifact against assertion-based tests. Any new compiler errors or runtime failures are fed back to the \textsc{Advisor Agent}. This loop continues until the IR passes LLVM verification and functional checks, or the iteration budget is exhausted.

\begin{figure}[t!]
 	\begin{center}
 		\includegraphics[width=0.49\textwidth]{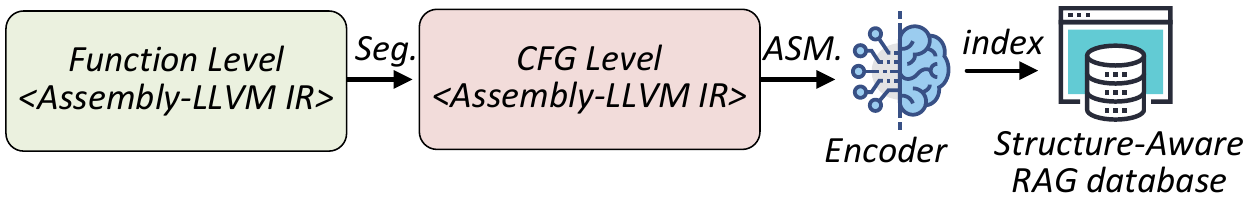}
 	\end{center}
  	\vspace{-3mm}
 	\caption{The process of building our Structure-Aware RAG database.}
 	 \vspace{-3mm}
 	\label{fig:rag}
 \end{figure}

 \section{Structure-Aware RAG Database\label{sec:rag}}

Figure~\ref{fig:rag} overviews the construction of a structure-aware RAG database used to lift optimized binaries to \emph{canonical} LLVM IR.

\subsection{Offline RAG Database Preparation}

We build an offline retrieval database from ExeBench~\cite{10.1145/3520312.3534867} to provide aligned exemplars for lifting machine code into pre-optimized, canonical LLVM IR. 

\cparagraph{Deduplication and leakage control.}
To improve diversity and avoid leakage, we (i) remove near-duplicate functions by computing normalized token-level edit similarity on canonical \texttt{-O0} LLVM IR and filtering pairs with similarity $>0.95$, and (ii) remove any samples that overlap with our evaluation set.

\cparagraph{Generating aligned assembly--IR pairs.}
For each retained program, we compile it to canonical LLVM IR using \texttt{clang -O0 -emit-llvm -S} as the ground-truth target.
We then compile the same source with Clang at four optimization levels (\texttt{-O0} to \texttt{-O3}) to obtain binaries, and disassemble each binary with \texttt{llvm-objdump} to produce the corresponding assembly.
After removing samples that fail to compile or disassemble, the final corpus contains 69,253 unique function-level canonical IR targets. Pairing each target with assembly from four optimization levels yields 277,012 function-level $(\langle \textit{assembly}, \textit{LLVM IR}\rangle)$ pairs.

\subsection{Control-Flow-Aware Data Structuring}

\begin{figure}[t]
 	\begin{center}
 		\includegraphics[width=0.5\textwidth]{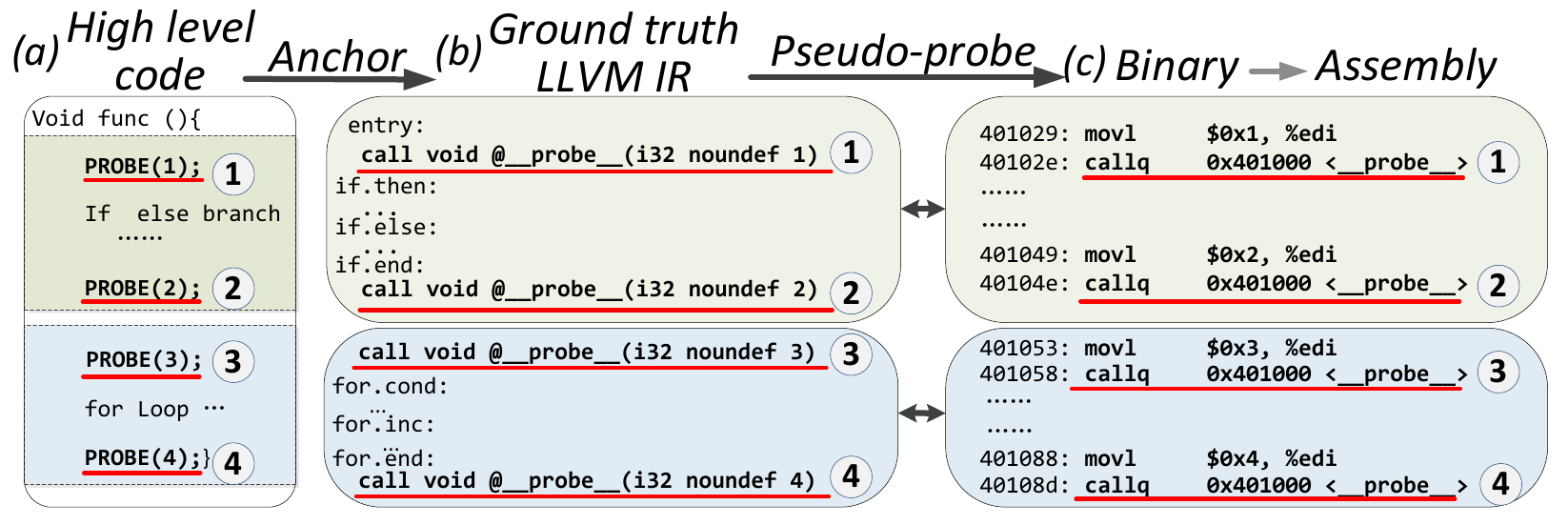}
 	\end{center}
  	\vspace{-4mm}
 	\caption{Workflow for constructing control-flow-aware LLVM IR and assembly code pairs.}
 	 \vspace{-3mm}
 	\label{fig:cfg-pairs}
 \end{figure}

Function-level pairs provide broad semantic coverage but bundle multiple control constructs into a single retrieval unit, making it difficult to isolate the structural patterns most relevant to a specific query region. We therefore construct \emph{region-level} entries that pair assembly snippets with structurally coherent IR slices at the granularity of individual control constructs (e.g., branch regions, loop bodies), providing a tighter alignment between retrieved exemplars and the target region at inference time. As illustrated in Figure~\ref{fig:cfg-pairs}.
Each region-level entry stores a triple $\langle \textit{asm}^{-Ox}, \textit{ir}^{-O0}, \textit{id}\rangle$ where $\textit{ir}^{-O0}$ is a canonical \texttt{-O0} IR slice and $\textit{asm}^{-Ox}$ is the corresponding assembly snippet at optimization level \texttt{-Ox}.

\cparagraph{(1) Identifying Structure region boundaries.}
We first identify boundaries of high-level control constructs in the source program.
Using an AST-based instrumentation script, we insert a lightweight marker \texttt{PROBE(x)} around constructs such as \texttt{if} conditionals and \texttt{for} loops.
and insert lightweight anchors \texttt{PROBE(x)} to mark region start/end points. These anchors are preserved when generating the canonical \texttt{-O0} LLVM IR(Figure~\ref{fig:cfg-pairs}a--b) and define structurally coherent regions for retrieval.

\cparagraph{(2) Fine-grained IR-to-assembly attribution with pseudo-probes.}
To propagate these semantic partitions to the final machine code, we leverage LLVM pseudo-probe instrumentation~\cite{pseudo_probe}. We inject lightweight pseudo-probes at IR basic-block entries so that probe identifiers are carried through optimization and emitted into the optimized binary. Pseudo-probes enable robust attribution of machine-instruction ranges back to their originating IR blocks, even when backend optimizations reorder instructions or rearrange basic blocks (Figure~\ref{fig:cfg-pairs}c). Consequently, we can precisely locate the high-level semantic boundaries in the optimized binary even in the presence of instruction scheduling or block
reordering.

\cparagraph{(3) Constructing structure-aware database entries.}
Finally, for each anchored region, we extract (i) the canonical \texttt{-O0} IR slice delimited by \texttt{PROBE} anchors and (ii) the corresponding region-complete assembly snippet obtained from pseudo-probe attribution.
These region-level pairs form the retrieval units used by \SystemName at inference time.
Across the corpus, this process yields 285K region-level entries.

\subsection{Embedding and Indexing}
We encode each segmented assembly region into a $d$-dimensional embedding using Nova-1.3B~\cite{jiang2025novagenerativelanguagemodels}.
Embeddings are stored as \texttt{float32} and indexed with FAISS using an exact \texttt{IndexFlatL2} wrapped by \texttt{IndexIDMap}~\cite{johnson2019billion}, enabling direct lookup of the paired $\langle \textit{assembly}, \textit{LLVM IR}\rangle$ entry via a unique region identifier.
Nova is chosen because it is trained to represent assembly patterns across optimization settings and can map diverse low-level realizations to similar semantics.
While the base encoder is strongest on x86-64, we adapt it to ARM64 and RISC-V with LoRA~\cite{hu2022lora} (rank $r{=}8$) using aligned region-level assembly pairs from our corpus, which improves cross-ISA retrieval robustness.

\section{Lift Agent\label{sec:retrieval}}
The \textsc{Lift Agent} produces an initial LLVM IR draft from an input binary.
It first disassembles the binary using \texttt{llvm-objdump}.
It then performs coarse control-flow recovery with \texttt{angr}~\cite{shoshitaishvili2016state} and applies a CFG-aligned segmentation procedure to partition the disassembly into control-structure regions (Section~\ref{sec:cfg-seg}).
Each region is used as a retrieval query to the previously mentioned structure-aware RAG database (Section~\ref{sec:rag}), which returns the top-$k$ aligned $\langle$\textit{assembly}, \textit{LLVM IR}$\rangle$ exemplars.
Conditioned on the query region and the retrieved exemplars, the agent constructs an in-context prompt and invokes the LLM to generate an initial LLVM IR candidate, which is then passed to the \textsc{Advisor Agent} for tool-guided refinement.

\subsection{CFG-aligned Segmentation}
\label{sec:cfg-seg}

\begin{figure}[t]
 	\begin{center}
 		\includegraphics[width=0.5\textwidth]{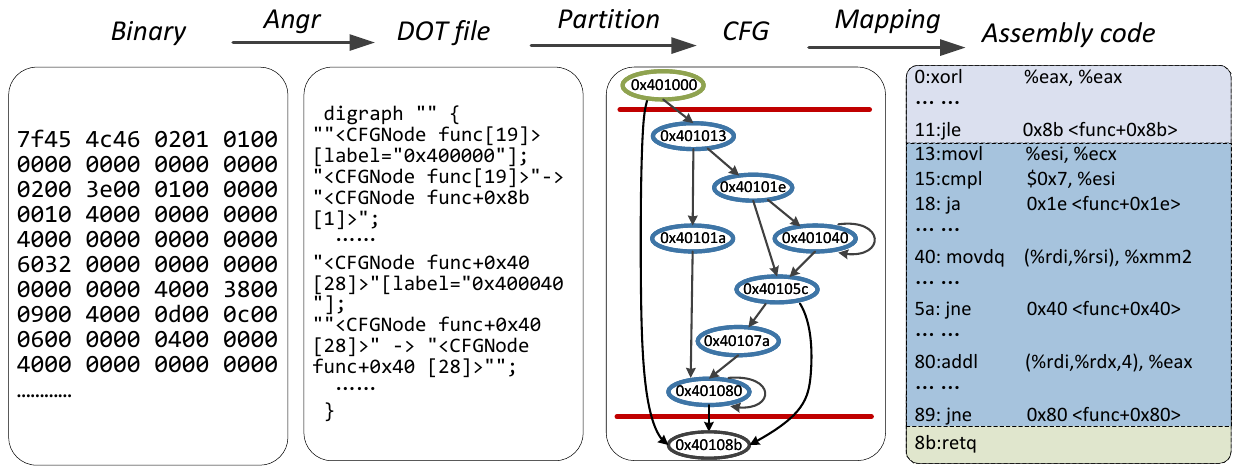}
 	\end{center}
  	\vspace{-3mm}
 	\caption{CFG-Guided binary segmentation.}
 	\label{fig:cfg-seg}
 \end{figure}

Figure~\ref{fig:cfg-seg} illustrates the segmentation procedure implemented as a tool that can be called by the \textsc{Lift Agent}.
Given a binary, we use \texttt{angr}~\cite{shoshitaishvili2016state} to recover a function-level CFG and export it in \texttt{DOT} format, where nodes correspond to basic blocks labeled by their start addresses.
We filter external stubs and unresolved callees (e.g., PLT entries) and segment only the internal control flow of the target function.

We identify candidate region headers at control-flow divergence points, defined as blocks with out-degree $\ge 2$.
Starting from each header, we follow successor edges along paths with out-degree $\le 1$ and aggregate visited blocks until reaching another divergence point or a function exit.
To handle loops, we detect back-edges during traversal and group the blocks on the cycle into a loop region.
For aggressively optimized code where unrolling yields multiple adjacent loop-like fragments, we merge contiguous loop fragments when they share a common header and rejoin pattern.

Finally, we merge regions corresponding to alternative branches when they converge at a common join block that is not a function exit.
The resulting regions provide stable retrieval units that capture local control structure (e.g., branch and loop regions) while remaining robust to CFG noise.
When CFG recovery is incomplete, we conservatively fall back to linear address-range regions derived from recovered basic blocks.

\subsection{Control-flow-aware Retrieval}
\label{sec:retrieval-tool}

Given a CFG-segmented assembly fragment, the \textsc{Lift Agent} computes a dense embedding using Nova-1.3B~\cite{jiang2025novagenerativelanguagemodels} and performs nearest-neighbor search over the FAISS index to retrieve the top-$k$ most similar exemplars. Each retrieved exemplar provides an aligned $\langle$\textit{assembly}, \textit{LLVM IR}$\rangle$ pair, which is used as in-context evidence for IR synthesis.
To quantify retrieval quality, we define retrieval accuracy as the fraction of queries for which at least one of the top-$k$ retrieved exemplars shares the same canonical region label as the query in the database. Under this metric, Nova-1.3B achieves Top-1/Top-3/Top-5 accuracy of 95.92\%, 98.61\%, and 98.83\%, respectively.

\subsection{Initial IR Synthesis}
\label{sec:lift-infer}
To synthesize an initial IR draft, the \textsc{Lift Agent} constructs an in-context prompt consisting of (i) the query assembly region and (ii) the top-$k$ retrieved exemplar pairs $\langle \textit{assembly}, \textit{LLVM IR}\rangle$, ordered by retrieval rank. The prompt is subject to a fixed context budget. To fit within the budget while preserving the target evidence, the agent may drop the lowest-ranked exemplar, while keeping the full query region intact. The final prompt instructs the LLM to emit a single-function LLVM IR draft in a canonical, pre-optimized style consistent with the target toolchain configuration. This draft is then passed to the \textsc{Advisor} and \textsc{Validation} agents for verification-guided refinement.

\begin{table}[t!]
    \centering
    \vspace{-1mm}
    \scriptsize
    \caption{Error types detected by \SystemName’s static analysis during LLVM IR verification.}
    \begin{tabular}{  p{1.5cm}| p{6cm} }
        \toprule
        \textbf{Error Type} & \textbf{Description} \\
        \midrule
        AssertError & This indicates that the function has failed its test cases within the expected time budget, reflecting a functional or behavioral error.\\
        \midrule
        FuncNameError & Arises when a function attempts to call an external function that is undeclared, or when a mismatch or conflict occurs in an external function's declaration.\\
        \midrule
        PhiError & A common SSA violation, typically caused by incorrect phi instruction usage. This includes referencing values not from immediate predecessor blocks or using operands with inconsistent types.\\
        \midrule
        Definition Conflict & Represents conflicts in variable definitions. This category encompasses errors such as TypeMismatch, UndefinedValueError, and MultiDefineError.\\
        
        \bottomrule
    \end{tabular}
    
    \vspace{-4mm}
    \label{tab:ir_error_types}
\end{table}

\section{Advisor Agent}
The \textsc{Advisor Agent} is responsible for error localization and repair planning given an initial lifted LLVM IR candidate. It aggregates diagnostics from the LLVM toolchain (e.g., \texttt{llvm-as} and \texttt{opt -verify}) and, when available, failures reported by the runtime test suite. It first assigns each failure to an error category and then localizes the failure to a concrete IR scope, such as a function, a basic block, or a small set of instructions. Concretely, it parses toolchain logs to extract the failing condition and referenced IR entities (SSA values, block labels, and type signatures), maps them to the corresponding IR lines, and constructs a minimal surrounding slice for context. We use failure categories to select specialized repair prompts and constraints, which reduces prompt noise and restricts edits to the minimal scope needed for the observed error, improving convergence and stability of the refinement loop. Based on the category and localized slice, the \textsc{Advisor Agent} produces a structured set of repair actions, including type adjustments, CFG consistency updates, insertion of missing declarations, or targeted rewrites of incorrect computations. These actions are passed to the \textsc{Validation Agent}, which applies them and re-runs compilation and runtime checks.

\begin{itemize}
\item \textbf{AssertError (behavioral mismatch).}
The IR passes LLVM validation but fails the assertion-based tests, indicating a semantic mismatch on the available inputs.  We derive a lightweight semantic summary from the disassembly (control-flow skeleton and key data dependencies) and provide the failing test evidence as an explicit behavioral signal. Conditioned on this structural summary and the observed test failure, the \textsc{Advisor Agent} prompts the LLM to propose targeted logic edits to the relevant IR slice.

\item \textbf{FuncNameError (linkage/ABI mismatch).}
This category covers unresolved external symbols, incorrect prototypes, and calling-convention mismatches. The agent extracts authoritative interface information from the binary (e.g., symbol table and relocation entries via \texttt{readelf}) and uses it to repair external declarations and call sites, ensuring that names, signatures, and calling conventions match the binary interface.

\item \textbf{PhiError (merge inconsistency at control-flow joins).}
These failures arise from invalid value merges at join points, such as missing incoming mappings, incompatible incoming types, or references to values not available along a predecessor path. The agent localizes the conflicting \texttt{phi} instructions, reconstructs the predecessor set for the join block, and provides explicit merge constraints (required incoming blocks and type consistency). The LLM is then prompted to repair the incoming mappings and types within the localized slice.

\item \textbf{DefinitionConflict (type and def--use inconsistency).}
These errors arise from conflicting definitions, type mismatches across uses, or inconsistent value propagation. The agent performs def--use slicing around the implicated values to collect local data-flow context from definitions to uses, and prompts the LLM to reconcile conflicting definitions and restore type consistency while preserving the surrounding computation.
\end{itemize}

\section{Validation Agent}
\label{sec:validation}

The \textsc{Validation Agent} applies the \textsc{Advisor Agent}'s repair actions to patch the initial LLVM IR draft and verifies the updated IR in an iterative loop with the \textsc{Advisor Agent}. Given a set of suggested edits, it first updates the IR and performs \emph{compilation verification} using the LLVM toolchain (e.g., \texttt{llvm-as} and \texttt{opt -verify}). If verification or compilation fails, the agent records the diagnostics together with the patched IR and returns them to the \textsc{Advisor Agent} for further localization and repair planning.

If compilation succeeds, the \textsc{Validation Agent} performs \emph{functional verification} by executing the compiled artifact under an assertion-based test suite. In practice, the dominant remaining failures are behavioral mismatches rather than toolchain errors. For a failing test, the agent extracts the relevant assembly slice and its recovered CFG, and prompts an LLM to summarize the function logic in natural language based on the CFG and the original disassembly. This summary serves as a semantic reference for subsequent repairs. The agent then packages the failure evidence, including the failing input, assertion message, crash trace when applicable, and the generated logic description, and returns it with the current IR to guide the next repair step.

The \textsc{Advisor} and \textsc{Validation} agents iterate until the IR passes toolchain verification and all available tests. To bound cost and ensure termination, we stop early on success and cap the loop at five repair iterations.

\section{Experimental Setup\label{sec:result-zxa}}

\subsection{Competitive Baselines}

We compare \SystemName with seven baselines, including two rule-based lifters, RetDec~\cite{retdec} and MCTOLL~\cite{10.1145/3316482.3326354}, and five general-purpose LLMs: DeepSeek-V3.2~\cite{deepseek_chat}, ChatGPT-4o~\cite{chatgpt2025}, Claude Sonnet 4.5~\cite{claude}, Qwen-Plus~\cite{qwen_website}, and Gemini-3~\cite{GoogleGemini}. MCTOLL supports only x86-64. All LLM baselines are accessed through their API interfaces. We evaluate each API-based baseline five times with the same prompt and report the average across runs.

We do not include LLM Compiler FTD~\cite{cummins2024metalargelanguagemodel} or Forklift~\cite{armengol-estape2024forklift}, because they target a different input setting: compiler-emitted assembly text (e.g., \texttt{clang -S}) rather than raw binary disassembly. Our task starts from \texttt{llvm-objdump} output extracted from optimized binaries, which differs substantially from compiler-emitted assembly due to stripped symbols, indirect control flow, and relocation-related artifacts.

\subsection{Evaluation Datasets}

We evaluate \SystemName and all baselines on HumanEval-Decompile and MBPP~\cite{austin2021program}. HumanEval-Decompile~\cite{tan2024llm4decompiledecompilingbinarycode} is a benchmark originally designed for binary-to-C decompilation and contains 164 programs with executable test cases. We adapt it to our setting by compiling the original C implementations into canonical LLVM IR.

We also use MBPP, a standard benchmark for code generation, and select 500 programs (IDs 11--510) that can be consistently compiled and executed in our pipeline. For both HumanEval-Decompile and MBPP, we first generate ground-truth canonical LLVM IR using a fixed Clang 19 \texttt{-O0} configuration, and then compile the same programs into binaries at four optimization levels (\texttt{-O0} to \texttt{-O3}). We extract assembly using \texttt{llvm-objdump} and use it as input to the lifting task.

\begin{table*}[t!]
\centering
	
  \scriptsize
  
    \caption{Comparison of binary-to-canonical LLVM IR lifting performance for \SystemName, MCTOLL, RetDec, and five general-purpose LLMs on HumanEval-Decompile and MBPP, across three ISAs and four optimization levels (\texttt{-O0} to \texttt{-O3}), measured by re-compilability, re-executability, and edit similarity. MCTOLL supports only x86-64, and RetDec supports x86-64 and ARM64. Green cells indicate the best result, and blue cells indicate the
second-best within each row and platform group.}
    \begin{tabular}{@{}ll|rrrrr|rrrrr|rrrrr@{}}
	  \toprule
		    \multicolumn{2}{c|}{\textbf{x86-64}} 	&\multicolumn{5}{c|}{Re-compilability (\%)}& \multicolumn{5}{c|}{Re-executability (\%)}& \multicolumn{5}{c}{Edit Similarity (\%)}\\
		\midrule
		Dataset&Opt. Level &-O0&-O1&-O2& -O3& Avg.&-O0&-O1&-O2&-O3 & Avg.& -O0& -O1& -O2& -O3 &Avg.\\

        \midrule
          \multirow{6}{*}{\shortstack{HumanEval-\\Decompile}}&MCTOLL& 87.2& 75.0& 63.5& 62.8& 72.1&\second{77.0}& \second{62.1}& \second{54.7}& \second{54.7}& \second{62.1}& 24.6& 22.0& 21.8& 21.6&22.5\\
           & RetDec&\second{93.9} &\second{92.7} &\second{92.1} &\second{91.5} &\second{92.5} &55.5 &48.2 &39.6 &37.8 &45.3 &35.6 &\second{35.5} &\second{32.0} &\second{31.7} &33.7\\
          &Qwen-plus& 25.7& 9.5& 8.7& 6.8& 12.7& 16.2& 9.4& 4.1& 3.4& 8.3 & 37.5& 23.6& 25.6& 25.4& 28.0 \\
          &Gemini-3 &42.7 &46.3 &45.1 &48.8 &45.7 &33.5 &30.5 &30.5 &26.2 &30.2 &30.7 &28.0 &26.7 &27.8 &28.3\\
          &ChatGPT-4o& 20.3& 7.4& 9.5& 8.8& 11.5& 11.5& 2.0& 2.0& 2.0& 4.4 & 41.8& 23.5& 25.0& 24.9& 28.8 \\
          &Claude Sonnet 4.5& 25.0& 27.4& 19.5& 22.3&18.3  & 18.3& 14.0& 8.5& 6.7&11.9  & 39.9& 28.7& 29.7& 29.9& 32.0 \\
          &DeepSeek-v3.2& 30.4& 35.8& 25.0& 22.3&28.4  & 14.2& 9.4& 7.4& 9.5&10.1  & \second{44.9}& 30.5& 30.7& 31.0& \second{34.3} \\
          &\textbf{\SystemName}& \best{100.0}& \best{99.4}& \best{98.8} &
          \best{98.8}&  \best{99.2}&  \best{96.3}& \best{95.1}& \best{92.7}& \best{93.9}& \best{94.5}& \best{64.3}& \best{47.0}& \best{46.6}& \best{47.7}& \best{51.4}\\
         \midrule
 \multirow{6}{*}{MBPP}& MCTOLL& 84.8& 76.2& 71.0& 70.4& 75.6& \second{74.2}& \second{65.0}&\second {60.4}& \second{ 59.8}& \second{64.9}& 25.5& 25.8& 22.2& 22.0&23.9\\
  & RetDec&\second{94.6} &\second{95.4} &\second{95.8} &\second{95.8} &\second{95.4} &49.4&48.0 &36.8 &36.4 &42.7 &34.2 &\second{34.0} &\second{30.0} &\second{29.7} &\second{32.0}\\
 & Qwen-plus& 36.2& 22.6& 18.6& 18.4& 24.0& 21.4& 8.8& 7.8& 7.8& 11.5& 33.6& 22.7& 23.9& 24.2& 26.1\\
 & Gemini-3 & 39.6& 43.4& 41.2& 41& 41.3& 36.0& 25.8& 27.8& 26.8& 29.1& 34.1& 28.6& 26.8& 24.1& 28.4\\
 & ChatGPT-4o& 28.0& 20.0& 18.0& 19.4& 21.4& 14.2& 7.6& 7.6& 7.8& 9.3& 38.5& 21.7& 22.4& 22.9& 26.4\\
 & Claude Sonnet 4.5& 40.8& 35.2& 33.2& 32.0& 35.3& 29.4& 21.2& 18.0& 17.4&21.5& \second{40.9}& 27.1& 28.6& 28.6& 31.3\\
 & DeepSeek-v3.2& 41.0& 35.2& 32.4& 31.0& 34.9& 25.0& 17.6& 14.2& 15.6& 18.1& 40.8& 25.5& 26.5& 26.6& 29.9\\
 & \textbf{\SystemName}& \best{98.8}& \best{98.8}& \best{98.2}& \best{98.8}& \best{98.7}& \best{96.2}& \best{95.0}& \best{94.6}& \best{95.8}& \best{95.4}& \best{61.7}& \best{46.8}& \best{46.3}& \best{46.0}& \best{50.2}\\
           
         \midrule
		  \multicolumn{2}{c|}{\textbf{ARM64}} 		&\multicolumn{5}{c|}{Re-compilability (\%)}& \multicolumn{5}{c|}{Re-executability (\%)}& \multicolumn{5}{c}{Edit Similarity (\%)}\\
               \midrule
          		Dataset&Opt. Level &-O0&-O1&-O2& -O3& Avg.&-O0&-O1&-O2&-O3 & Avg.& -O0& -O1& -O2& -O3 &Avg.\\
           
         \midrule
          \multirow{6}{*}{\shortstack{HumanEval-\\Decompile}}&RetDec&\second{93.9} &\second{92.1} &\second{92.7} &\second{92.7} &\second{92.8} &0 &0.6 &0.6 &0.6 &0.5  &33.8  &\second{34.0}&\second{32.1}&\second {31.8}& \second{32.9}\\
          &Qwen-plus&26.2 &11.1 &18.5 &13.6 &17.4 &8.5 &0.6 &1.2 &2.5 &3.2  &35.2  &25.1 & 25.7 &25.9 &28.0  \\
          &Gemini-3 &29.5 &43.9 &45.1&46.3 & 41.1 &\second{23.2} & \second{35.4}&\second{34.1} &\second{23.2}&\second{29.0} & 35.0& 32.7 &31.3 &30.4 & 32.4\\
          &ChatGPT-4o&20.1 &11.7 &9.3 &10.5 &12.9  & 8.5&3.1 & 1.2& 1.2&3.5  &31.3 &23.0  &23.6 & 23.5 &25.4  \\
         & Claude Sonnet 4.5&10.4 &26.2 &20.1 &22.0 &19.7 & 7.9 &11.0 & 7.9 & 7.9 &8.7 &38.5 &27.5 &27.9 & 27.5 &30.4\\
          &DeepSeek-v3.2&27.4 &42.6 &34.0 &35.5 &34.6 &13.4 &10.5 &7.4 &8.6 &10.0&\second{41.9}&27.7 &28.3 &28.2 &31.6 \\
          &\textbf{\SystemName}&\best{98.2} &\best{99.4} &\best{98.1} &\best{99.4} &\best{98.8} &\best{94.5} &\best{95.1} &\best{92.6} &\best{94.4} &\best{94.2} &\best{60.8} &\best{39.0} &\best{38.6} &\best{38.5} &\best{44.3} \\
         \midrule
 \multirow{6}{*}{MBPP}& RetDec&\second{96.0} &\second{93.4} &\second{93.4} &\second{93.2} &\second{94.0} &0.2 &0.4 &0.4 &0.4 &0.4  &32.2  &\second{31.8}& \second{29.8}& \second{29.4} &30.7\\ 
 & Qwen-plus&36.0 &25.3 &23.0 &22.5 &26.7 &19.6 &9.4 &8.6 &8.2 &11.5 &33.8 &24.1 &24.4 &25.0 &26.9 \\
 & Gemini-3 &31.4 &40.0 &39.2 &36.6 &36.8 &\second{25.0} &\second{30.6} &\second{29.8} &\second{28.6} &\second{28.5} &39.0 &29.5 &\second{29.8} &\second{29.4} &\second{31.9} \\
 & ChatGPT-4o&33.2 &18.2 &16.2 &15.6 &21.4 &18.2 &7.8 &7.0 &6.8 &10.0 &30.6 &22.0 &22.2 &22.4 &24.3 \\
 & Claude Sonnet 4.5&16.8 &22.2 &22.4 &22.8 &21.1 &12.0 &11.6 &11.0 &11.0 &11.4 &38.9 &25.5 &26.2 &26.6 &29.3\\
 & DeepSeek-v3.2&39.2 &46.4 &44.6 &42.3 &43.1  &17.6 &15.9 &15.7 &15.5 &16.2   &\second{40.2} &24.5  &25.5 &25.3  &29.0  \\
 & \textbf{\SystemName}&\best{96.4} &\best{97.6} &\best{96.1} &\best{97.3} &\best{96.6} &\best{92.6} &\best{93.7} &\best{90.4} &\best{91.4} &\best{92.0} &\best{58.1} &\best{37.4} &\best{37.5} &\best{37.7} &\best{42.8} \\
          \midrule
		  \multicolumn{2}{c|}{\textbf{RISC-V}} 		&\multicolumn{5}{c|}{Re-compilability (\%)}& \multicolumn{5}{c|}{Re-executability (\%)}& \multicolumn{5}{c}{Edit Similarity (\%)}\\
               \midrule
          		Dataset&Opt. Level &-O0&-O1&-O2& -O3& Avg.&-O0&-O1&-O2&-O3 & Avg.& -O0& -O1& -O2& -O3 &Avg.\\
                            \midrule
         \multirow{6}{*}{\shortstack{HumanEval-\\Decompile}}&Qwen-plus&42.1&25.9&29.0&32.1&32.3&5.5&0.6&1.9&1.9&2.4&34.1&20.8&20.0&20.7 &23.9\\
          &Gemini-3 &52.4&\second{55.6}&\second{66.1}&\second{71.6}&\second{61.4} &\second{36.0}&\second{22.2} &\second{25.3}&\second{26.5}&\second{27.5}&\second{42.0} &\second{35.2} &\second{35.2}&\second{35.1}&\second{36.9} \\
          &ChatGPT-4o&20.1&10.5&7.4&8.0&11.5&6.1 &1.9&1.9&1.9 &2.9&28.1 &22.3&22.6&22.6&23.9\\
         & Claude Sonnet 4.5&17.7&25.9&26.5&24.7&23.7&8.5 &9.9&11.7 &8.6 &9.7&28.8&22.9&22.7&22.6 &24.3\\
          &DeepSeek-v3.2&\second{56.7}&48.8&43.8&53.7&50.8&12.8&9.9&8.0&12.3&10.8&40.6&28.5&28.3&28.4&31.5\\
          &\textbf{\SystemName}&\best{99.4}&\best{97.5}&\best{96.9}&\best{97.5}&\best{97.8}&\best{88.4}&\best{79.0}&\best{77.8}&\best{77.2}&\best{80.6}&\best{63.9}&\best{48.1}&\best{48.0}&\best{47.2}&\best{51.8}\\ 
               \midrule
  \multirow{6}{*}{MBPP}& Qwen-plus&52.8&32.2&33.5&31.4&37.6&9.4&2.7&3.3&3.3&4.7&29.8&19.8&19.6&19.6&22.2\\
 & Gemini-3 &64.6&\second{69.8}&\second{72.7}&\second{71.8}&\second{69.7}&\second{41.8}&\second{33.1}&\second{30.1}&\second{29.2}&\second{33.8}&\second{39.2}&\second{33.2}&\second{33.2}&\second{33.2}&\second{34.7}\\
 & ChatGPT-4o&28.8&17.1&20.8&18.0&21.2&9.8&6.1&7.6&6.1&7.4&28.0&20.9&20.6&21.2&22.6\\
 & Claude Sonnet 4.5&37.6&43.1&39.6&41.8&40.5&21.2&18.0&16.7&17.1&18.3&28.6&20.4&20.9&20.8&22.7\\
 & DeepSeek-v3.2&\second{65.6}&59.8&58.4&57.6&60.4&17.0&13.7&14.9&14.9&15.2&38.1&26.4&26.0&26.4&29.2\\
 & \textbf{\SystemName}&\best{99.2}&\best{98.8} &\best {98.2}&\best{99.4}&\best{98.9}&\best{91.8}&\best{88.8}&\best{89.8}&\best{89.8}&\best{90.0}&\best{46.3}&\best{40.0}&\best{39.7}&\best{39.0}&\best{41.3}\\
   \bottomrule
	\end{tabular}
         	
\label{tab:result-decompile}
		 \vspace{-5mm}
\end{table*}

\vspace{-1mm}
\subsection{Metrics}
We use five metrics to evaluate lifting quality: re-compilability, re-executability, edit similarity, CFG full-match accuracy, and CFG similarity. These five metrics serve as the primary evaluation measures for lifting performance~\cite{wong2025decllm,tan2024llm4decompiledecompilingbinarycode,tan2025sk2decompile}, capturing toolchain validity, behavioral correctness, and structural fidelity to the canonical reference IR.  We also evaluate \emph{analysis-friendliness} and \emph{migration ability} on lifted IR separately in Section~\ref{sec:downstream} and Section~\ref{sec:migration}.

\textbf{Re-compilability} measures the percentage of lifted LLVM IR functions that can be successfully assembled and compiled without errors.

\textbf{Re-executability} measures the percentage of lifted functions that not only compile successfully but also pass the associated test cases, reflecting behavioral consistency with the reference program under the available inputs.

\textbf{Edit similarity}~\cite{armengolestape2024sladeportablesmalllanguage} measures the similarity between the lifted LLVM IR and the ground-truth canonical IR using normalized edit distance. Higher values indicate closer correspondence to the reference IR.

\textbf{CFG full-match accuracy} measures the percentage of functions whose reconstructed control-flow graphs are exactly identical to the reference CFG. We compute this metric using \texttt{networkx.isomorphic(G1, G2)}~\cite{SciPyProceedings_11}, which tests whether the two graphs are isomorphic under an edge-preserving one-to-one node mapping.

\textbf{CFG similarity} quantifies structural similarity between the lifted and reference CFGs using the Weisfeiler--Lehman graph kernel~\cite{JMLR:v21:18-370}. Higher values indicate closer agreement in graph structure even when the CFGs are not exactly identical.

\section{Experimental Results\label{sec:result}}
\subsection{Overall Performance}

Table~\ref{tab:result-decompile} summarizes the overall lifting performance of \SystemName and seven baselines across two benchmarks, three ISAs, and four compiler optimization levels.

We can observe that \SystemName  achieves the highest re-executability in nearly all reported settings. On HumanEval-Decompile, it reaches 94.5\%  on x86-64, 94.2\% on ARM64, and 80.6\% on RISC-V (averaged over \texttt{-O0}--\texttt{-O3}). The strongest non-\SystemName baselines remain below 30\% on ARM64 and RISC-V, indicating that direct lifting with retrieval and refinement is substantially more effective at preserving executable semantics across architectures than open-loop generation.

Across both benchmarks, we observe a clear separation between re-compilability and re-executability. Rule-based lifters, especially RetDec, often achieve high re-compilability (e.g., 92.5\% on x86-64 and 92.8\% on ARM64 for HumanEval-Decompile), yet their re-executability is much lower (45.3\% on x86-64 and 0.5\% on ARM64). A similar gap appears on MBPP, where RetDec remains highly compilable (95.4\% on x86-64 and 94.0\% on ARM64) but achieves limited re-executability (42.7\% and 0.4\%). This pattern shows that producing verifier-accepted IR is comparatively easy, while preserving behavior from optimized binaries is the dominant challenge.

\SystemName also produces IR that is closer to the fixed canonical \texttt{-O0} reference. On HumanEval-Decompile, its average edit similarity is 51.0\% (x86-64), 44.3\% (ARM64), and 51.8\% (RISC-V), exceeding all baselines in the corresponding settings. The same trend holds on MBPP, where \SystemName improves edit similarity while achieving substantially higher re-executability. Taken together, these results suggest that improved executability is not achieved by unconstrained rewriting, but by converging toward a representation closer to the canonical target.

In detail, on x86-64, \SystemName achieves the best performance on all three metrics across both HumanEval-Decompile and MBPP. On ARM64, it again outperforms RetDec on all metrics for both benchmarks. This shows that retrieval-augmented lifting with refinement improves both semantic preservation and syntactic correctness.
On RISC-V, where no rule-based baselines are available, \SystemName delivers the largest margin over all general-purpose LLMs, reaching 97.8\% re-compilability and 80.6\% re-executability on HumanEval-Decompile. These gains indicate that the framework generalizes well beyond x86-style control-flow and instruction patterns.

The optimization-level breakdown shows that \SystemName is also robust under aggressive compiler optimization. On HumanEval-Decompile, the re-compilability of \SystemName remains above 98\% both on x86-64 and on ARM64 from \texttt{-O0} to \texttt{-O3}, while re-executability remains comparatively stable. On MBPP, \SystemName also maintains stable re-executability across four optimization levels, consistently outperforming all baselines.

\begin{figure}[t]
    \subfloat[][CFG Accuracy (\%)]{\includegraphics[width=0.24\textwidth]{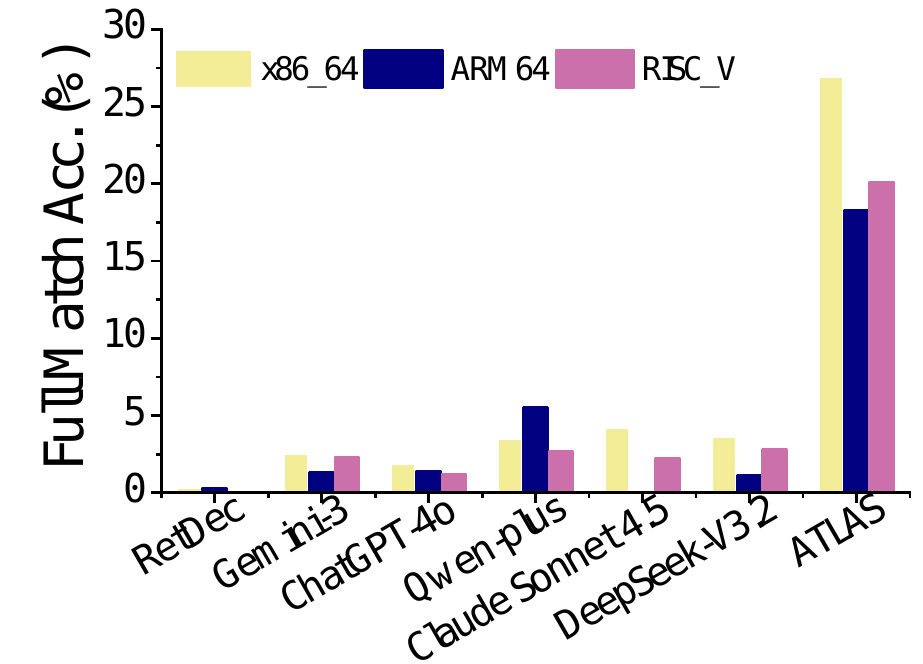}}
	\subfloat[][CFG Similarity (\%)]{\includegraphics[width=0.24\textwidth]{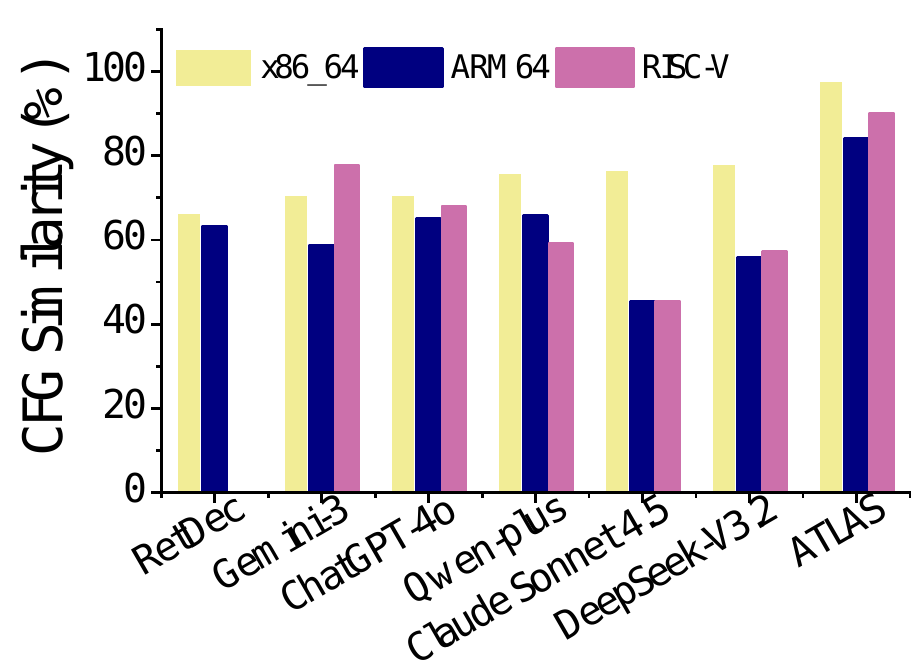}}
    \caption{LLVM IR CFG reconstruction performance on the HumanEval-Decompile datasets across x86-64, ARM64 and RISC-V.} 	\vspace{-4mm}
 	\label{fig:cfg-sim}
 \end{figure}

Figure~\ref{fig:cfg-sim} evaluates CFG reconstruction using two metrics: full-match CFG accuracy (graph isomorphism) and CFG similarity (WL-kernel similarity, normalized to $[0,100]$). Exact CFG recovery is challenging under optimization because transformations such as if-conversion, block merging/splitting, and indirect control transfers can substantially reshape the graph, and accordingly all baselines achieve low full-match accuracy (below 6\% on x86-64, ARM64, and RISC-V). In contrast, \SystemName attains 26.9\%, 18.3\%, and 20.1\% full-match accuracy on the three ISAs, indicating a markedly stronger ability to recover complete control-flow structure rather than only partial fragments.CFG similarity shows the same pattern: \SystemName reaches 97.8, 84.1, and 90.2, consistently outperforming all baselines and demonstrating that its reconstructed CFGs remain close to the reference even when exact isomorphism is not achieved. The RISC-V results are particularly informative: although full-match accuracy is lower than on x86-64/ARM64, CFG similarity remains high, suggesting that \SystemName recovers most high-level control-flow organization while some fine-grained details remain difficult under aggressive optimization. Overall, these results support the effectiveness of \SystemName’s region-level retrieval and verification-guided refinement in recovering compiler-meaningful control-flow structure across architectures.

\begin{table*}[!t]
\centering

  \vspace{-1mm}
  \scriptsize
\caption{Initial LLVM IR quality produced by the Lift Agent with RAG under different retrieval granularities on HumanEval-Decompile (x86-64).}
  \vspace{-2mm}
    \begin{tabular}{@{}l|rrrrr|rrrrr|rrrrr@{}}
	  \toprule
		  &\multicolumn{5}{c|}{Re-compilability (\%)}& \multicolumn{5}{c|}{Re-executability (\%)}& \multicolumn{5}{c}{Edit Similarity (\%)}\\
		\midrule
		Opt. Level&-O0&-O1&-O2& -O3& Avg.&-O0&-O1&-O2&-O3 & Avg.& -O0& -O1& -O2& -O3 &Avg.\\
        \midrule
         Basic Block level& 85.5& 71.6& 63.4& 53.2& 68.4 & 66.2 & 48.5 & 41.9& 40.2&49.2 & 58.8& 40.5& 40.2& 39.8& 44.8 \\
         Control flow level& \second{90.2}& \best{86.6}& \best{84.1}& \best{90.2}& \best{87.8} & \second{67.7} & \second{50.6 }& \best{53.7}& \best{55.5} & \best{56.9} & \second{63.8} & \best{46.6 }& \best{46.2}& \best{47.3}& \best{51.0}\\
         Function level&\best{91.2} & \second{83.6}& \second{82.4} & \second{79.4}& \second{84.1} & \best{70.6} & \best{52.9} & \second{52.9} & \second{41.1} & \second{54.4}& \best{65.0}& \second{40.0}& \second{42.5}& \second{40.0}& \second{46.9} \\

           \bottomrule
	\end{tabular}

\label{tab:multi-levels-rag}
\end{table*}

\begin{table*}[!t]
\centering
           \caption{Effect of the \textsc{Advisor}--\textsc{Validation} refinement loop on HumanEval-Decompile (x86-64).
Without refinement loop: \textsc{Lift}-only. With refinement loop: full \SystemName.}
  \vspace{-1mm}
  \scriptsize
    \begin{tabular}{@{}l|rrrrr|rrrrr|rrrrr@{}}
	  \toprule
		  &\multicolumn{5}{c|}{Re-compilability (\%)}& \multicolumn{5}{c|}{Re-executability (\%)}& \multicolumn{5}{c}{Edit Similarity (\%)}\\
		\midrule
		Opt. Level&-O0&-O1&-O2& -O3& Avg.&-O0&-O1&-O2&-O3 & Avg.& -O0& -O1& -O2& -O3 &Avg.\\
        \midrule
         Without refinement loop& 90.2& 86.6& 84.1& 90.2& 87.8 & 67.7 & 50.6 & 53.7& 55.5 & 56.9 & 63.8 & 46.6 & 46.2& 47.3& 51.0\\
         With refinement loop& \best{100.0}& \best{99.4}& \best{98.8} & \best{98.8}& \best{99.2}& \best{96.3}& \best{95.1}& \best{92.7}& \best{93.9}& \best{94.5}& \best{64.3}& \best{47.0} &\best{46.6}& \best{47.7}& \best{51.4}\\
           \bottomrule
	\end{tabular}
	
\label{tab:refinement-loop}
		 \vspace{-4mm}
\end{table*}

\subsection{Multi-Level RAG in Lift Agent}
To study how retrieval granularity affects LLVM IR reconstruction, we compare three RAG configurations that retrieve exemplars at different code granularities: basic-block level, control-flow (region) level, and function level. Basic-block retrieval emphasizes local instruction patterns, control-flow retrieval captures branch and loop structure, and function-level retrieval provides the broadest semantic context. To isolate the effect of retrieval granularity, we disable the iterative refinement loop and perform a single inference pass for each setting.

Table~\ref{tab:multi-levels-rag} shows that \textbf{control-flow–level retrieval provides the strongest overall balance} across all three metrics. It achieves the highest average re-compilability (87.8\%), the highest average re-executability (56.9\%), and the highest average edit similarity (51.0\%). Compared with basic-block retrieval, control-flow retrieval improves average re-compilability by 19.4 points (68.4\%$\rightarrow$87.8\%) and average re-executability by 7.7 points (49.2\%$\rightarrow$56.9\%), indicating that conditioning on region-level structure is important for generating verifier-clean IR and preserving executable semantics. The advantage is most visible under heavier optimizations, where basic-block retrieval degrades substantially at \texttt{-O3} (53.2\% re-compilability and 40.2\% re-executability), while control-flow retrieval remains stable (90.2\% re-compilability and 55.5\% re-executability). This suggests that region-level exemplars provide structural cues that remain informative even when local instruction sequences are heavily rewritten by optimization.

Function-level retrieval improves on basic-block retrieval but does not outperform control-flow retrieval overall. While function-level retrieval achieves the best re-compilability at \texttt{-O0} (91.2\%) and the best re-executability at \texttt{-O0} (70.6\%), its performance degrades more at higher optimization levels, especially for re-executability at \texttt{-O3} (41.1\%). A key reason is that aggressive optimizations (e.g., inlining, if-conversion, and code motion) substantially alter whole-function structure and instruction composition, making function-level similarity less discriminative for the specific region that determines behavior. As a result, whole-function retrieval is more likely to return exemplars that are globally similar but locally mismatched, which can bias generation toward an incorrect computation while still producing compilable IR. In contrast, control-flow–level retrieval matches exemplars at the same granularity used during inference and repair, providing region-aligned structural cues that remain more stable under optimization and improving semantic fidelity.

\subsection{Ablation: Impact of the Advisor--Validation Refinement Loop}
\label{sec:ablation_loop}

Table~\ref{tab:refinement-loop} isolates the contribution of the iterative \textsc{Advisor}--\textsc{Validation} loop. The \textbf{Lift-only} setting reports the initial IR produced by the \textsc{Lift Agent} (single-pass synthesis with the same segmentation and retrieval), without invoking the \textsc{Advisor} or \textsc{Validation} agents. The \textbf{full} setting runs \SystemName with the complete refinement loop.

Enabling the refinement loop improves both toolchain validity and functional correctness. Re-compilability increases from 87.8\% to 99.2\% (+11.4 points), while re-executability rises from 56.9\% to 94.5\% (+37.6 points). The gains are consistent across optimization levels, with particularly large improvements at \texttt{-O1}--\texttt{-O3}, where optimized binaries introduce more structural and semantic ambiguity that is difficult to resolve in a single pass.

Edit similarity remains nearly unchanged (51.0\% vs.\ 51.4\%), suggesting that refinement primarily performs localized, diagnostic-driven repairs rather than unconstrained rewriting. Overall, this ablation shows that retrieval-grounded initial synthesis provides a strong starting point, but the tool-guided, closed-loop refinement is the main driver of end-to-end correctness.

\begin{table*}[t!]
\centering
\footnotesize
\caption{SVF-based downstream analyzability on HumanEval-Decompile (x86-64).}
\setlength{\tabcolsep}{6pt}
\renewcommand{\arraystretch}{1.15}
\begin{tabularx}{\linewidth}{lccc>{\raggedright\arraybackslash}X}
\toprule
\textbf{Metric} & \textbf{LLM4Decompile+clang} & \textbf{\SystemName} & \textbf{RetDec} & \textbf{Description} \\
\midrule
SVF completion (\%) $\uparrow$ & 81.7 & \best{85.5} & 82.5 & Fraction of functions for which SVF finishes successfully under a fixed configuration and resource budget. Higher values indicate that the lifted IR is more analysis-ready. \\
Avg. alias set size $\downarrow$ & 1.05 & \best{1.00} & 1.03 & Mean size of SVF-computed alias sets. Smaller values indicate more precise pointer reasoning. \\
Pointer Coverage Index (PCI) $\uparrow$ & 80.4 & \best{132.9} & 107.1 & Normalized measure of how much pointer and value-flow structure in the lifted IR can be recovered by SVF. Higher values indicate richer downstream analyzability. \\
\bottomrule
\end{tabularx}
\vspace{2pt}

\label{tab:svf}
\end{table*}

\begin{table}[t]
\centering
\footnotesize
\setlength{\tabcolsep}{6pt}
\caption{Cross-ISA migration on HumanEval-Decompile (x86-64 $\rightarrow$ ARM64). Each method lifts an x86-64 binary to LLVM IR, then recompiles the lifted IR for ARM64. Re-compilability and re-executability are measured on the ARM64 target.}
\label{tab:crossisa_compare}
\begin{tabular}{lcc}
\toprule
\textbf{Method} & \textbf{Re-compilability} & \textbf{ Re-executability} \\
\midrule
RetDec         & 90.6\% & 44.1\% \\
LLM4Decompile  & 82.5\% & 40.7\%  \\
\SystemName    & 98.3\% & 93.6\% \\
\bottomrule
\end{tabular}
\end{table}

\section{Downstream Analyzability}
\label{sec:downstream}

A key motivation for lifting binaries to canonical LLVM IR is to support downstream compiler and security analyses when source code is unavailable. In many real-world settings, such as third-party software auditing, legacy-system maintenance, and binary-only migration, the deployed binary is the only trustworthy artifact. Lifting such binaries to LLVM IR enables the reuse of mature compiler and static-analysis infrastructures while remaining closer to deployed behavior than analyses performed on reconstructed high-level code. However, compilability alone does not imply that the lifted IR is suitable for downstream analysis. Rule-based lifters may produce executable IR that is structurally fragile or overly type-erased, while decompilation-to-C pipelines primarily optimize for readability and executability rather than analysis-oriented IR quality. Our goal is therefore not only to recover executable code, but to reconstruct pre-optimized, canonical LLVM IR that is verifier-valid, analyzable, and reusable for downstream workflows.

We evaluate whether the lifted IR can support downstream pointer and value-flow analysis using SVF. We compare \SystemName with a Binary$\rightarrow$C$\rightarrow$IR baseline, \textbf{LLM4Decompile+clang}, on HumanEval-Decompile. We report three metrics. \textbf{SVF completion rate} measures the fraction of functions on which SVF finishes successfully under a fixed configuration and resource budget. \textbf{Average alias set size} measures the mean size of SVF-computed alias sets, where smaller values indicate more precise pointer reasoning. \textbf{Pointer Coverage Index (PCI)} measures how much pointer-related structure in the lifted IR can be successfully recovered and tracked by SVF. Higher PCI indicates that the IR preserves clearer pointer types and memory flows, enabling SVF to construct richer pointer/value-flow graphs.

Table~\ref{tab:svf} shows that \SystemName produces LLVM IR that is more amenable to downstream analysis than both the Binary$\rightarrow$C$\rightarrow$IR baseline and the rule-based lifter RetDec. SVF completes on 85.5\% of functions for \SystemName, compared with 81.7\% for LLM4Decompile+clang and 82.5\% for RetDec, indicating that \SystemName yields IR that is more consistently analyzable under the same tool configuration. For functions that SVF can process, \SystemName also achieves the smallest average alias set size (1.00), suggesting slightly better pointer precision than the baselines. The largest gain is observed in PCI, where \SystemName reaches 132.9, substantially higher than LLM4Decompile+clang (80.4) and RetDec (107.1). The results indicate that direct lifting to canonical LLVM IR better preserves analysis-relevant typing and memory-flow information, improving both SVF robustness and the informativeness of downstream value-flow graphs.

\section{Downstream Cross-ISA Migration}
\label{sec:migration}

To evaluate whether \SystemName recovers architecture-independent program semantics rather than ISA-specific surface patterns, we study a downstream \emph{cross-ISA migration} task. The key idea is that canonical, pre-optimization LLVM IR decouples high-level program logic from the source instruction set, making it possible to recompile lifted binaries for a different target architecture. This provides a direct test of whether the recovered IR can serve as a reusable intermediate representation for software migration.

We first compare \SystemName with representative baselines for cross-ISA migration from x86-64 to ARM64. All methods start from an x86-64 binary and transform it into high-level code, which is then recompiled for ARM64 and evaluated on the ARM64 target. \textbf{RetDec} follows a rule-based lifting pipeline that translates the x86-64 binary into LLVM IR using static recovery heuristics. \textbf{LLM4Decompile} follows a Binary to C pipeline, where the x86-64 binary is first decompiled to C and then recompiled for ARM64 platform. \SystemName directly lifts the x86-64 binary to canonical LLVM IR and refines it with tool-guided feedback. 
As shown in Table~\ref{tab:crossisa_compare}, \SystemName achieves the best migration performance, reaching 98.3\% re-compilability and 93.6\% re-executability after recompilation to ARM64. In contrast, RetDec attains 90.6\% re-compilability but only 44.1\% re-executability, while LLM4Decompile yields 82.5\% re-compilability and 40.7\% re-executability. The large gap in re-executability indicates that \SystemName more reliably preserves architecture-independent semantics required for cross-target reuse, whereas the baselines often produce artifacts that recompile but fail to preserve behavior under migration.

\begin{table}
\centering
\footnotesize
\caption{Cross-target \textbf{re-compilability} of \SystemName on HumanEval-Decompile. Rows denote the source ISA of the lifted binary, and columns denote the target ISA used for recompilation.}
\label{tab:recompilability}
\begin{tabular}{lccc}
\toprule
\textbf{Source ISA} & \textbf{Target x86-64} & \textbf{Target ARM64} & \textbf{Target RISC-V} \\
\midrule
\textbf{x86-64}  & --& 98.3\% & 98.3\% \\
\textbf{ARM64}   &98.8\% & --& 98.8\% \\
\textbf{RISC-V } & 97.8\% & 97.8\% & --\\
\bottomrule
\end{tabular}
\end{table}

\begin{table}
\centering
\footnotesize
\caption{Cross-target \textbf{ re-executability} of \SystemName on HumanEval-Decompile. Rows denote the source ISA of the lifted binary, and columns denote the target ISA used for recompilation and execution.}
\label{tab:reexecutability}
\begin{tabular}{lccc}
\toprule
\textbf{Source ISA} & \textbf{Target x86-64} & \textbf{Target ARM64} & \textbf{Target RISC-V} \\
\midrule
\textbf{x86-64}  & --& 93.6\% & 93.6\% \\
\textbf{ARM64 }  & 94.2\% & --& 94.2\% \\
\textbf{RISC-V } & 80.0\% & 80.5\% & --\\
\bottomrule
\end{tabular}
\end{table}

We further report full source--target transfer matrices in Tables~\ref{tab:recompilability} and~\ref{tab:reexecutability}. 
Table~\ref{tab:recompilability} shows that \SystemName maintains high recompilation success across all source--target pairs, with all results exceeding 97\%. Table~\ref{tab:reexecutability} demonstrates that these recompilation gains translate to functional success, with the best cross-ISA migration results again obtained when lifting from x86-64 and recompiling to ARM64 or RISC-V (both 93.6\%). Similarly, transfers from ARM64 and RISC-V also exhibit strong performance. Overall, these results indicate that \SystemName recovers a stable canonical IR that supports cross-architecture recompilation and execution. The consistently strong transfer performance further suggests that \SystemName is not merely translating local instruction patterns, but reconstructing an architecture-independent intermediate representation that preserves program logic across targets, which is essential for migration workflows in binary-only settings.

\section{Further Analysis}

\begin{table*}[!t]
\centering
\caption{\textbf{Portability across backbone LLMs.} Performance on HumanEval-Decompile when using a general-purpose LLM alone (\textbf{Base}) versus augmenting it with our framework (\textbf{+ATLAS}: structure-aware RAG + Advisor--Validation refinement).}
\vspace{-2mm}
\rowcolors{2}{gray!10}{white}
\scriptsize
\begin{tabular}{l|rrrrr|rrrrr|rrrrr}
\toprule
&\multicolumn{5}{c|}{Re-compilability (\%)}& \multicolumn{5}{c|}{Re-executability (\%)}& \multicolumn{5}{c}{Edit Similarity (\%)}\\
\midrule
\textbf{Model / Setting} &-O0&-O1&-O2&-O3&Avg.&-O0&-O1&-O2&-O3&Avg.&-O0&-O1&-O2&-O3&Avg.\\
\midrule
Qwen-Plus (Base) & 25.7& 9.5& 8.7& 6.8& 12.7& 16.2& 9.4& 4.1& 3.4& 8.3& 37.5& 23.6& 25.6& 25.4& 28.0 \\
Qwen-Plus (+ATLAS) & 69.6& 58.1& 54.1& 50.0& 58.0& 51.4& 40.5& 38.5& 37.8& 42.1& 54.4& 40.6& 39.8& 40.9& 43.9 \\

Claude Sonnet 4.5 (Base) & 25.0& 27.4& 19.5& 22.3& 18.3& 18.3& 14.0& 8.5& 6.7& 11.9& 39.9& 28.7& 29.7& 29.9& 32.0 \\
Claude Sonnet 4.5 (+ATLAS) & 70.1& 60.4& 54.3& 53.0& 59.5& 50.6& 47.6& 37.8& 35.4& 42.8& 49.7& 40.5& 41.5& 49.5& 45.3 \\

ChatGPT-4o (Base) & 20.3& 7.4& 9.5& 8.8& 11.5& 11.5& 2.0& 2.0& 2.0& 4.4& 41.8& 23.5& 25.0& 24.9& 28.8 \\
ChatGPT-4o (+ATLAS) & 57.4& 42.6& 45.3& 43.2& 48.6& 42.6& 37.2& 35.8& 35.8& 37.9& 50.4& 41.1& 41.8& 41.3& 43.7 \\

DeepSeek-V3.2 (Base) & 30.4& 35.8& 25.0& 22.3& 28.4& 14.2& 9.4& 7.4& 9.5& 10.1& 44.9& 30.5& 30.7& 31.0& 34.3 \\
DeepSeek-V3.2 (+ATLAS) & \second{94.6}& \second{90.5}& \second{83.8}& \second{83.8}& \second{88.2}& \second{86.6}& \second{67.1}& \second{61.0}& \second{57.9}& \second{68.2}& \second{63.6}& \second{41.3}& \second{43.5}& \second{43.4}& \second{47.9} \\

Gemini-3 (Base) & 42.7& 46.3& 45.1& 48.8& 45.7& 33.5& 30.5& 30.5& 26.2& 30.2& 30.7& 28.0& 26.7& 27.8& 28.3 \\
ATLAS (Gemini-3 backbone) & \best{100.0}& \best{99.4}& \best{98.8}& \best{98.8}& \best{99.2}& \best{96.3}& \best{95.1}& \best{92.7}& \best{93.9}& \best{94.5}& \best{64.3}& \best{47.0}& \best{46.6}& \best{47.7}& \best{51.4} \\
\bottomrule
\end{tabular}
\label{tab:re-opt-levels}
\end{table*}

\subsection{Generalization Across Backbone LLMs}
\label{sec:backbone}

To isolate the contribution of our lifting pipeline from the choice of backbone model, we instantiate the framework with five general-purpose LLMs: Qwen-Plus, Claude Sonnet 4.5, ChatGPT-4o, DeepSeek-V3.2, and Gemini-3. Table~\ref{tab:re-opt-levels} reports results on HumanEval-Decompile across four optimization levels (\texttt{-O0}--\texttt{-O3}). For each backbone, we compare the base model against the same model augmented with our framework (structure-aware retrieval and the Advisor--Validation refinement loop), using identical prompts, the same retrieval database, and the same refinement settings.

Table~\ref{tab:re-opt-levels} shows that the \SystemName consistently improves all backbone models on all three primary lifting metrics. Averaged across the five backbones, our method increases re-compilability from 24.4\% to 70.7\%, re-executability from 13.0\% to 54.3\%, and edit similarity from 30.3\% to 46.4\%. These gains indicate that the benefit does not come from a single backbone, but from the combination of control-flow-aware retrieval and verification-guided refinement.

The greatest improvements are observed for DeepSeek-V3.2 and Gemini-3. For DeepSeek-V3.2, our scheme improves average re-compilability from 28.4\% to 88.2\% and re-executability from 10.1\% to 68.2\%. For Gemini-3, the same scheme raises re-compilability from 45.7\% to 99.2\% and re-executability from 30.2\% to 94.5\%, while also improving edit similarity from 28.3\% to 51.4\%. This result suggests that stronger base models benefit most from the structured retrieval and repair loop, since they can better exploit retrieved exemplars and tool feedback.

The improvements are also consistent across optimization levels. For weaker backbones such as Qwen-Plus and ChatGPT-4o, the raw models degrade sharply as optimization increases, but the proposed scheme substantially narrows this gap. For example, Qwen-Plus improves recompilability from 6.8\% to 50.0\% and re-executability from 3.4\% to 37.9\% at \texttt{-O3}. In contrast, \SystemName remains stable across optimization levels, maintaining 98.8--100.0\% re-compilability and 93.9--96.1\% re-executability from \texttt{-O0} to \texttt{-O3}. Overall, these results show that the proposed framework is backbone-agnostic and that its gains arise primarily from the structure-aware RAG and iterative refinement design, rather than from relying on a particular proprietary model.

\begin{figure}[t!]
 	\begin{center}
 		\includegraphics[width=0.4\textwidth]{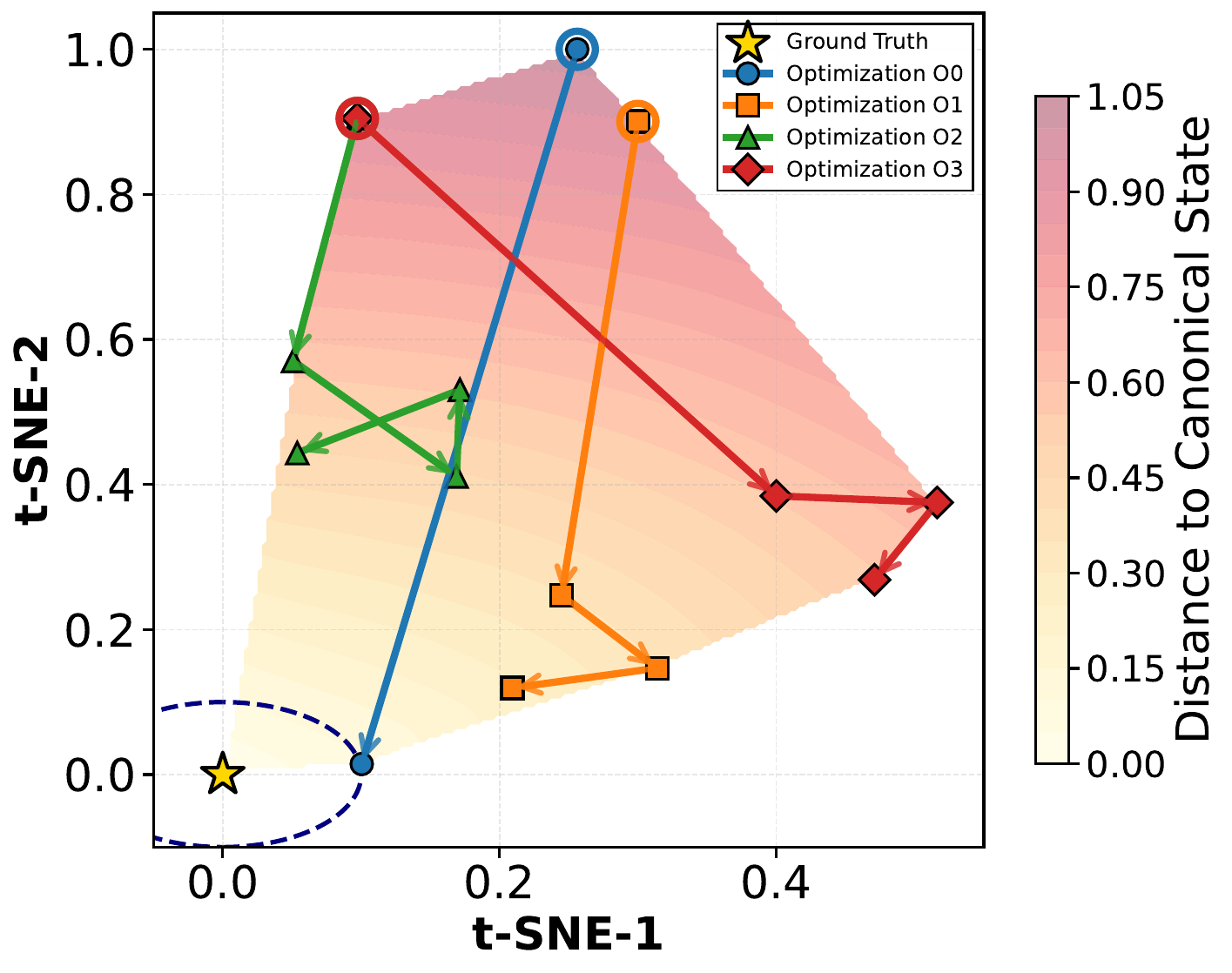}
 	\end{center}
  	\vspace{-3mm}
 	\caption{t-SNE visualization of the semantic refinement trajectory for one test case (ID:132) in the HumanEval-Decompile dataset. We project the latent embeddings of the lifted LLVM IR using \texttt{IR2vector}~\cite{VenkataKeerthy-2020-IR2Vec}. The paths trace the evolution from the initial input through five iterative refinement cycles toward the Ground Truth (Red Star).}
 	\label{fig:tsne}
 \end{figure}

\subsection{Lifting Direction Correction }
Figure~\ref{fig:tsne} provides a qualitative view of how iterative refinement changes the lifted IR in representation space. We embed each intermediate LLVM IR produced during refinement using IR2Vec~\cite{VenkataKeerthy-2020-IR2Vec} and project the embeddings to two dimensions with t-SNE. Each trajectory starts from the initial draft (here initialized from the rule-based lifter MCTOLL) and follows successive refinement steps toward the ground-truth canonical IR.
Across optimization levels, the initial drafts lie far from the ground-truth embedding, suggesting that rule-based lifting often produces IR that is structurally and operationally different from the canonical reference. For \texttt{-O0} inputs, the trajectory typically reaches the neighborhood of the ground truth in one refinement step, consistent with the lower degree of optimization-induced rewriting.  In contrast, \texttt{-O1}--\texttt{-O3} inputs typically require multiple iterations, with trajectories moving toward the reference in a stepwise manner. We can also observe a two-phase behavior: the largest displacement toward the ground truth often occurs during the initial synthesis stage, indicating that retrieval-grounded generation provides coarse alignment to the canonical \texttt{-O0} target even for highly optimized binaries. Subsequent refinement steps make smaller, localized movements, consistent with targeted repairs that address remaining legality and behavioral mismatches while staying within the same semantic neighborhood. Although t-SNE is qualitative and does not by itself establish semantic equivalence, this pattern is consistent with our ablations showing that retrieval substantially improves the initial draft and that the Advisor--Validation loop yields additional gains in end-to-end re-executability.

\begin{table}[!t]
\centering

  \small
    \caption{Impact of the iterative refinement count on \SystemName performance on the HumanEval-Decompile dataset (x86-64). $k=0$ represents the initial inference without feedback.}
    \begin{tabular}{@{}c|ccc}
	  \toprule
		  \textbf{Iteration}&\textbf{ Re-Comp (\%)}&\textbf{ Re-Exec (\%)}& \textbf{Edit-Sim (\%)}\\
        \midrule
         0& 87.8 &56.9 & 50.9 \\
         1& 94.8 &78.4 & 51.2 \\
         2& 97.3 &88.1 & 51.3 \\
         3& 97.9 &91.9 & 51.4 \\
         4& 98.9 &93.3 & 51.4 \\
         5& 99.2 &94.5 & 51.4 \\

 6&98.8 &94.8 & 51.4\\
 7&99.5 &95.0 & 51.4\\
             \bottomrule
	\end{tabular}
 		\vspace{-2mm}

\label{tab:refinement-iterations}
		 \vspace{-4mm}
\end{table}

\subsection{Impact of iterative refinement.}
Table~\ref{tab:refinement-iterations} evaluates the effect of the refinement-loop iteration count ($k$). The results show that iterative refinement substantially improves both re-compilability and re-executability in the early stages. Starting from the initial draft without feedback ($k=0$), re-compilability increases from 87.8\% to 99.2\% at $k=5$, while re-executability improves from 56.9\% to 94.5\%, a gain of 37.6 percentage points. These results indicate that the feedback loop is effective at correcting both structural and behavioral errors in the lifted IR. In contrast, edit similarity remains essentially unchanged throughout refinement (50.9\% at $k=0$ and 51.4\% from $k=1$ onward), suggesting that the iterative process improves functional correctness and verifier validity without materially changing the overall textual proximity to the canonical reference IR. Beyond $k=5$, the gains become marginal: re-executability increases only slightly from 94.5\% to 95.0\% at $k=7$, while re-compilability no longer improves and even drops slightly at $k=7$. Therefore, we set $k=5$ as the default iteration budget, since it captures most of the benefit of refinement while avoiding unnecessary computational overhead.

\begin{figure}[t]
        \begin{center}
 		\includegraphics[width=0.3\textwidth]{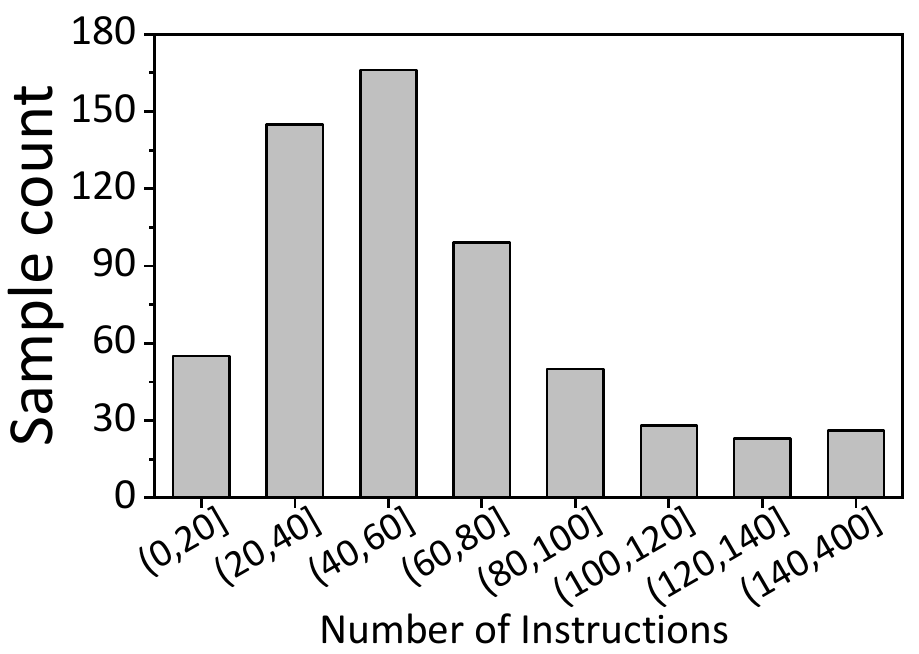}
 	\end{center}
  	\vspace{-4mm}
 	\caption{Distribution of input assembly lengths (measured by the number of instructions) in the HumanEval-Decompile-IR benchmarks.}
 	 \vspace{-5mm}
 	\label{fig:number}
 \end{figure}

 \begin{figure*}[t!]
    \centering
    \subfloat[][Re-compilability (\%)]{\includegraphics[width=0.29\textwidth]{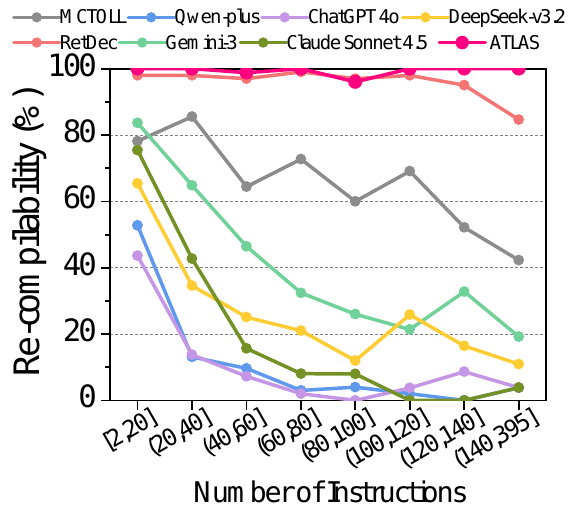}}
	\subfloat[][Re-executability (\%)]{\includegraphics[width=0.29\textwidth]{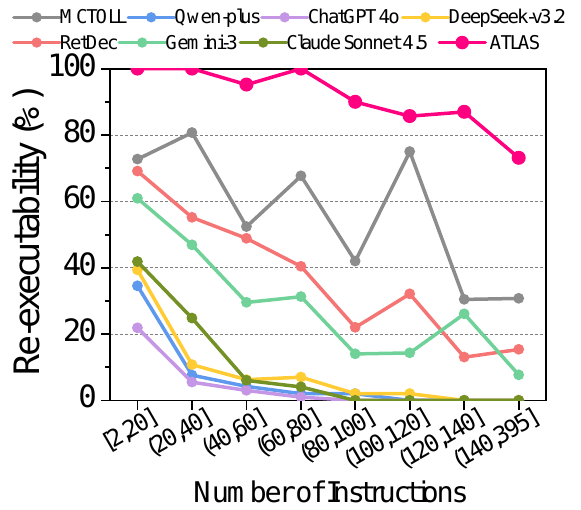}}
    \subfloat[][Edit Similarity (\%)]{\includegraphics[width=0.29\textwidth]{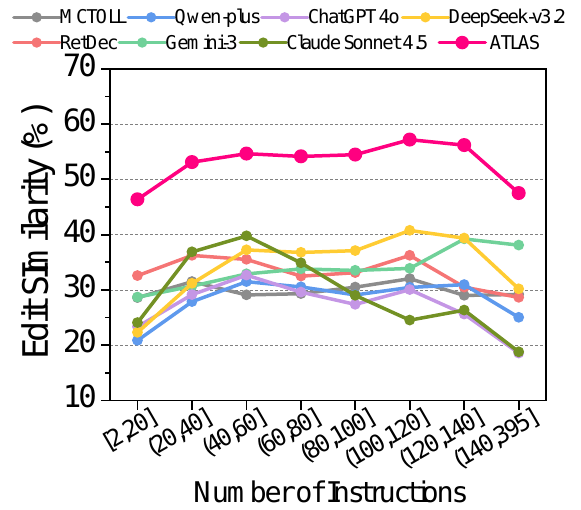}}\\
     \subfloat[][Node Accuracy (\%)]{\includegraphics[width=0.29\textwidth]{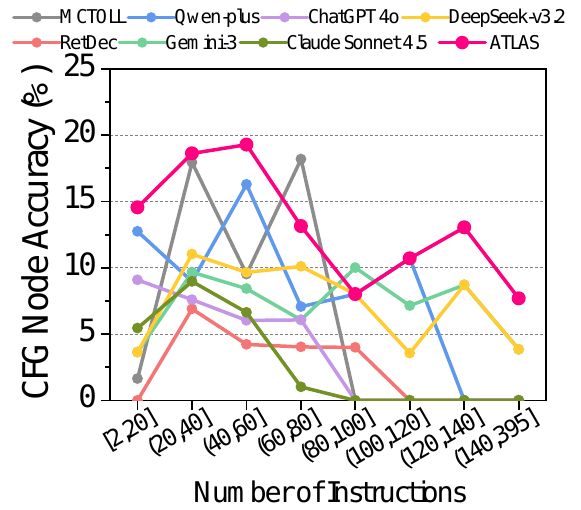}}
	\subfloat[][CFG Full Match Accuracy (\%)]{\includegraphics[width=0.29\textwidth]{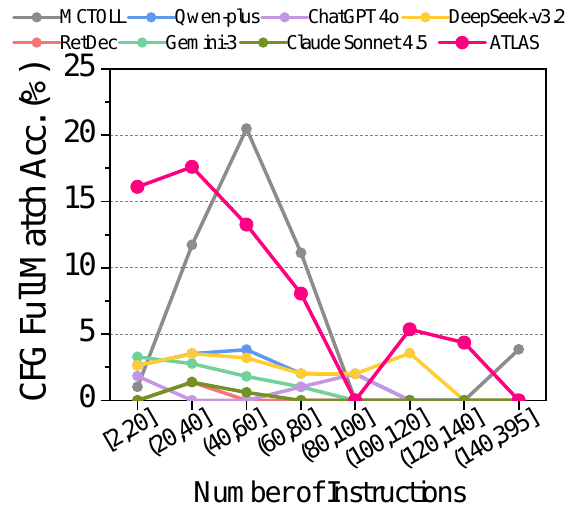}}
    \subfloat[][CFG Similarity (\%)]{\includegraphics[width=0.29\textwidth]{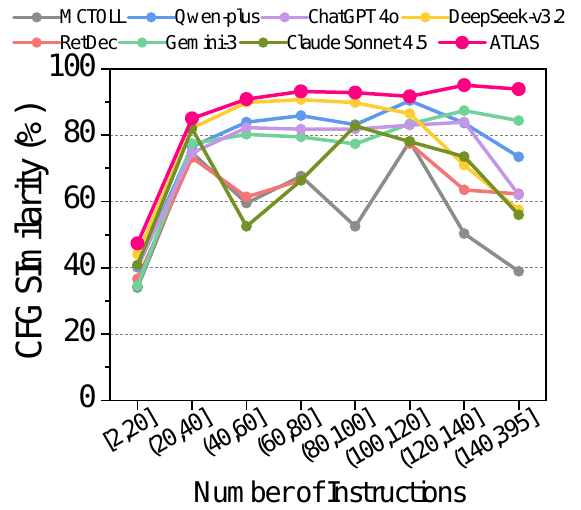}}
 	\caption{\SystemName lifting performance comparison across different input assembly code lengths of HumanEval-Decompile dataset (x86-64).}
 	 \vspace{-5mm}
 	\label{fig:inst-perf}
 \end{figure*}

 \begin{figure*}[t!]
    \centering
    \subfloat[][Re-compilability (\%)]{\includegraphics[width=0.29\textwidth]{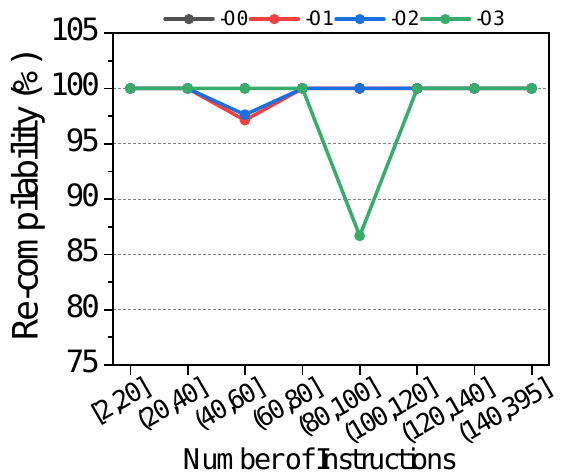}}
	\subfloat[][Re-executability (\%)]{\includegraphics[width=0.29\textwidth]{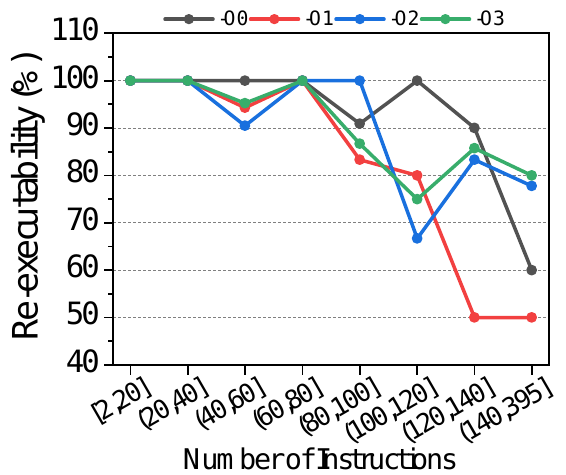}}
    \subfloat[][Edit Similarity (\%)]{\includegraphics[width=0.29\textwidth]{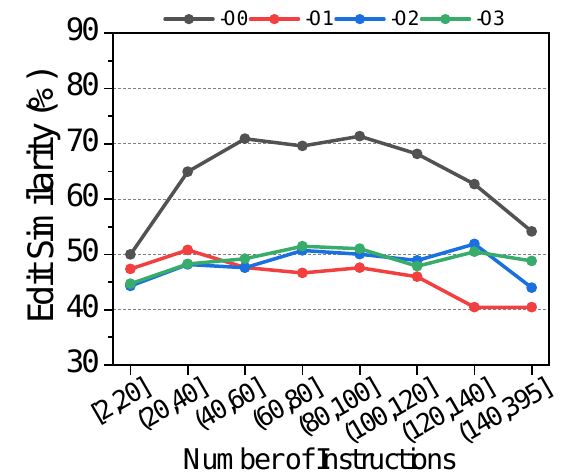}}\\
     \subfloat[][Node Accuracy (\%)]{\includegraphics[width=0.29\textwidth]{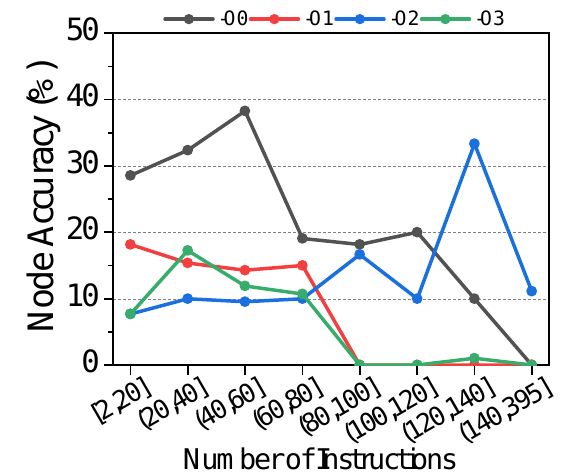}}
	\subfloat[][CFG Full Match Accuracy (\%)]{\includegraphics[width=0.29\textwidth]{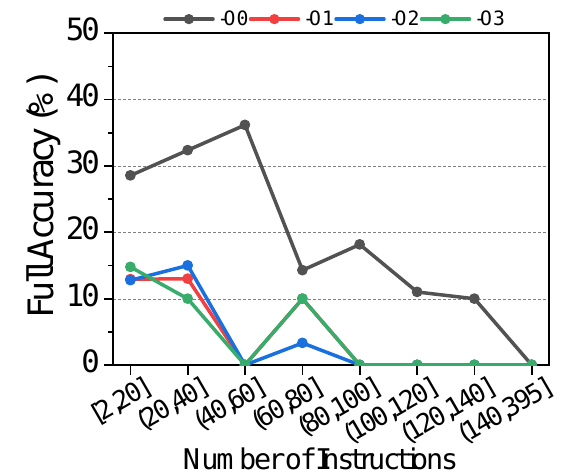}}
    \subfloat[][CFG Similarity (\%)]{\includegraphics[width=0.29\textwidth]{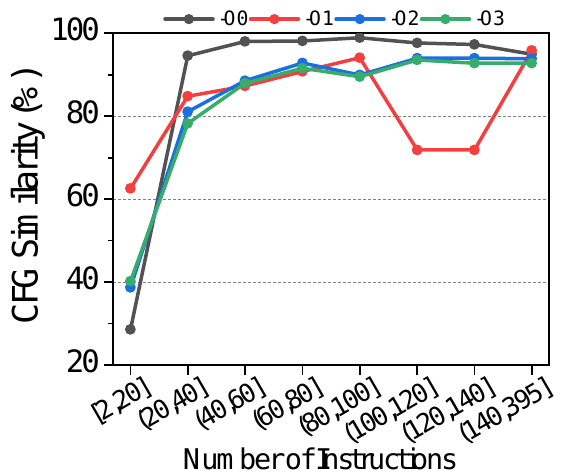}}
 	\caption{\SystemName's performance across varying input assembly code lengths and optimization levels on the HumanEval-Decompile dataset (x86-64).}
 	 \vspace{-3mm}
 	\label{fig:opt-overall}
 \end{figure*}

\subsection{Impact of Input Length on Decompilation Performance}

We also investigate the impact of assembly code length on the lifting performance of \SystemName. Figure~\ref{fig:number} illustrates the distribution of function lengths in the HumanEval-Decompile dataset, where the majority concentrate in the 40–60 (166 functions) and 20–40 (145 functions) instruction ranges. Figure~\ref{fig:inst-perf} compares \SystemName with seven baselines across six evaluation metrics: re-compilability, re-executability, edit similarity, node accuracy (which measures whether the node count matches the ground truth), CFG full-match accuracy, and CFG similarity, covering input lengths from 2 to 394 instructions (approximately 42 to 8,325 tokens). 
 
We can observe that general-purpose LLMs exhibit a sharp performance degradation as function complexity increases. In the range of [2, 20] instructions, general-purpose LLMs achieve an average re-executability of 39.7\%. When the instruction count exceeds 60, re-executability drops below 9\% for all models except Gemini-3, which retains a higher rate of 31.0\%, making them ineffective for non-trivial functions.
In contrast, \SystemName significantly extends this effective window. It maintains robust functional correctness well into medium-length sequences, achieving 100\% re-executability at [60, 80] instructions, whereas ChatGPT-4o exhibits the lowest re-executability among all general-purpose LLMs at 1.0\%, demonstrating that structure-aware retrieval effectively mitigates the context erosion that typically hampers standard LLMs.  Even in the longest sequences ([140, 395] instructions), \SystemName retains 100\% Re-compilability and 93.8\% CFG Similarity. This significantly outperforms the best LLM baselines, proving that our framework preserves the global topology of massive functions where others produce non-compilable code. While structural recovery remains high, Re-executability drops to 73.1\% for the longest functions. This divergence suggests that while \SystemName successfully recovers the "program skeleton," the accumulation of minor logic errors in very long sequences remains a challenge for end-to-end execution.

Figure~\ref{fig:opt-overall} presents the lifting performance of \SystemName across varying optimization levels and input instruction counts. We can observe that the re-executability remains high for all levels in short sequences, but -O1 and -O3 exhibit a decline at the extreme long-tail (140+ instructions). This trend indicates that aggressive transformations, such as inlining and loop restructuring, introduce functional complexities that challenge the \SystemName in long-context scenarios. 
Regarding textual fidelity, the \SystemName maintains a robust average edit similarity of 64.3\% for \texttt{-O0}. However, this metric decreased in the long-tail regime (140-395 instructions, accounting for 54.2\% of the HumanEval-Decompile). Moreover, it is worth noting that in the $[2, 20]$ instruction range, we observe a decoupling of functional and structural metrics. While \SystemName achieves near-perfect \textit{Re-executability}, both \textit{CFG Similarity} (28.6\%) and \textit{Edit Similarity} (49.0\%) remain low. This is not an indication of lifting failure, but rather a reflection of metric sensitivity to sparse data. In small functions, minor variations in register naming, function attributes, or the inclusion of entry/exit blocks represent a significant percentage of the total code or graph structure, leading to disproportionately low similarity scores despite functional equivalence.

Table~\ref{tab:longfunc} lists the performance of \SystemName on a harder subset of long functions, including 56 cases with more than 200 instructions from HumanEval and MBPP.  We compare \SystemName with the rule-based lifter RetDec on this subset. 
RetDec achieves relatively high re-compilability (91.1\%), yet its re-executability is only 10.7\% and edit similarity is 23.9\%, indicating that the generated code often compiles but fails to preserve the ground-truth program behavior. In contrast, \SystemName achieves 100\% re-compilability on these long functions, while substantially improving re-executability to 76.9\% and increasing edit similarity to 44.4\%. This demonstrates that \SystemName not only preserves structural correctness but also significantly enhances functional fidelity, even for structurally complex functions. These results highlight that \SystemName scales effectively to long functions, maintaining both syntactic validity and semantic consistency, whereas rule-based approaches tend to degrade in behavioral correctness under increased structural complexity.

\begin{figure*}[t!]
    \centering
    \subfloat[][\SystemName one shot inference]{\includegraphics[width=0.23\textwidth]{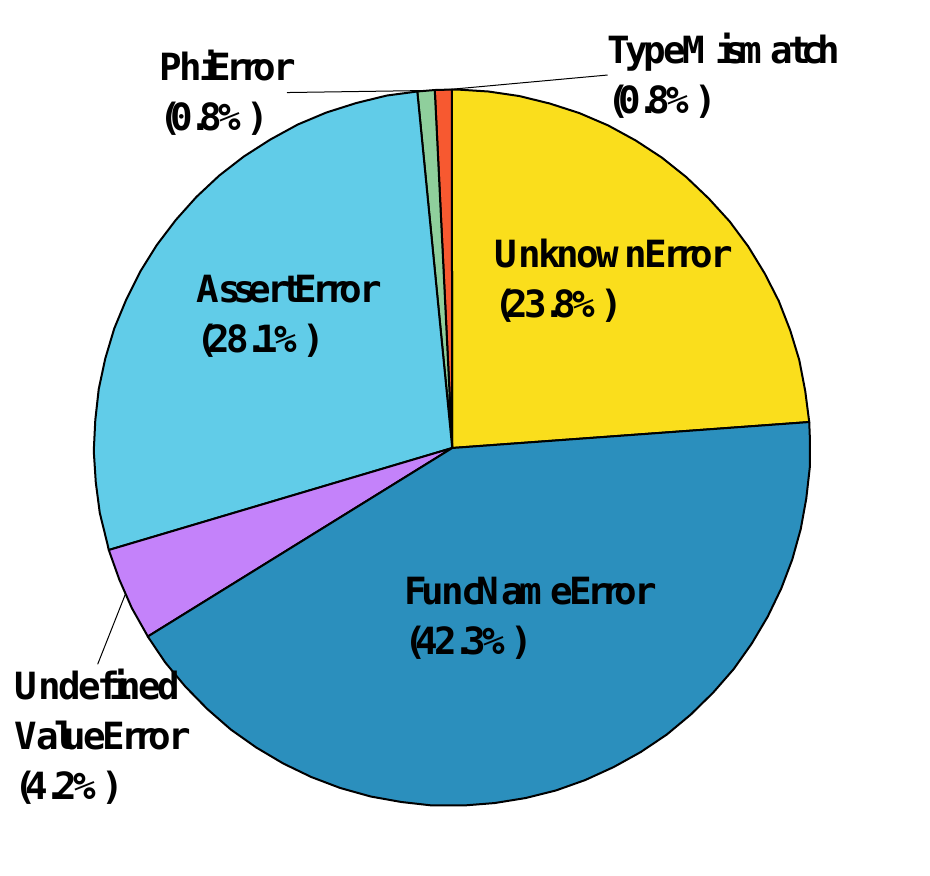}}
	\subfloat[][\SystemName with one-pass feedback refinement]
{\includegraphics[width=0.23\textwidth]{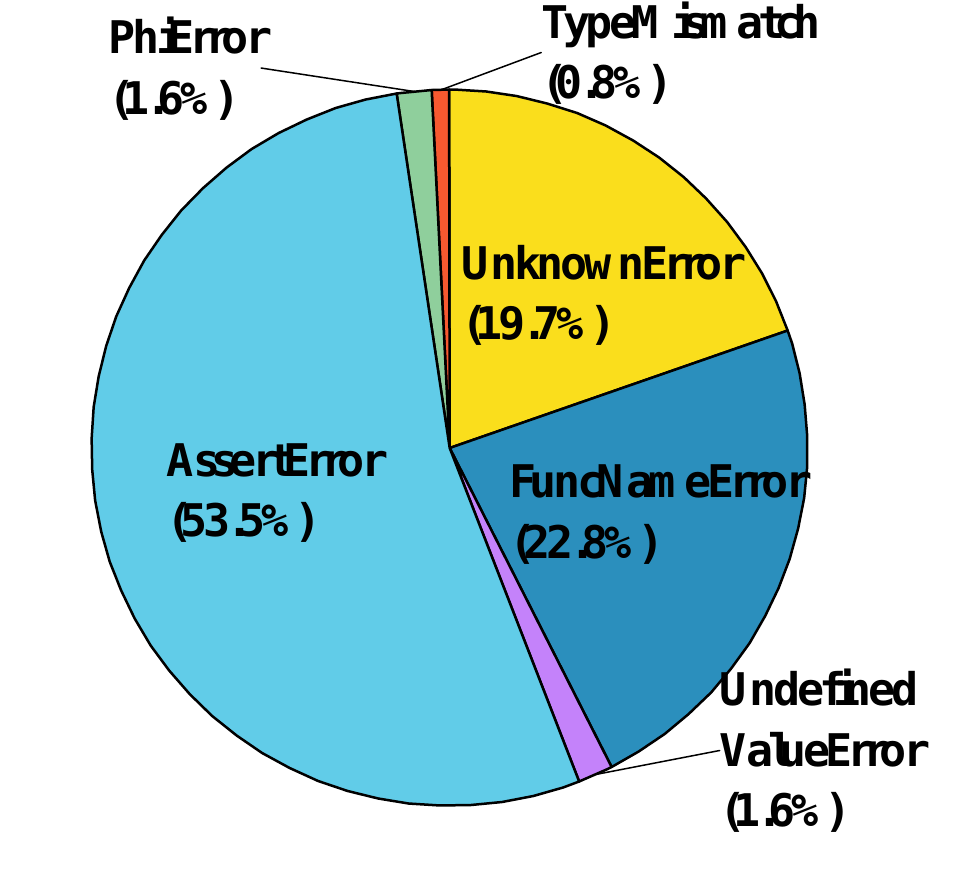}}
    \hspace{2mm}
    \subfloat[][\SystemName  with three-pass feedback refinement.]{\includegraphics[width=0.23\textwidth]{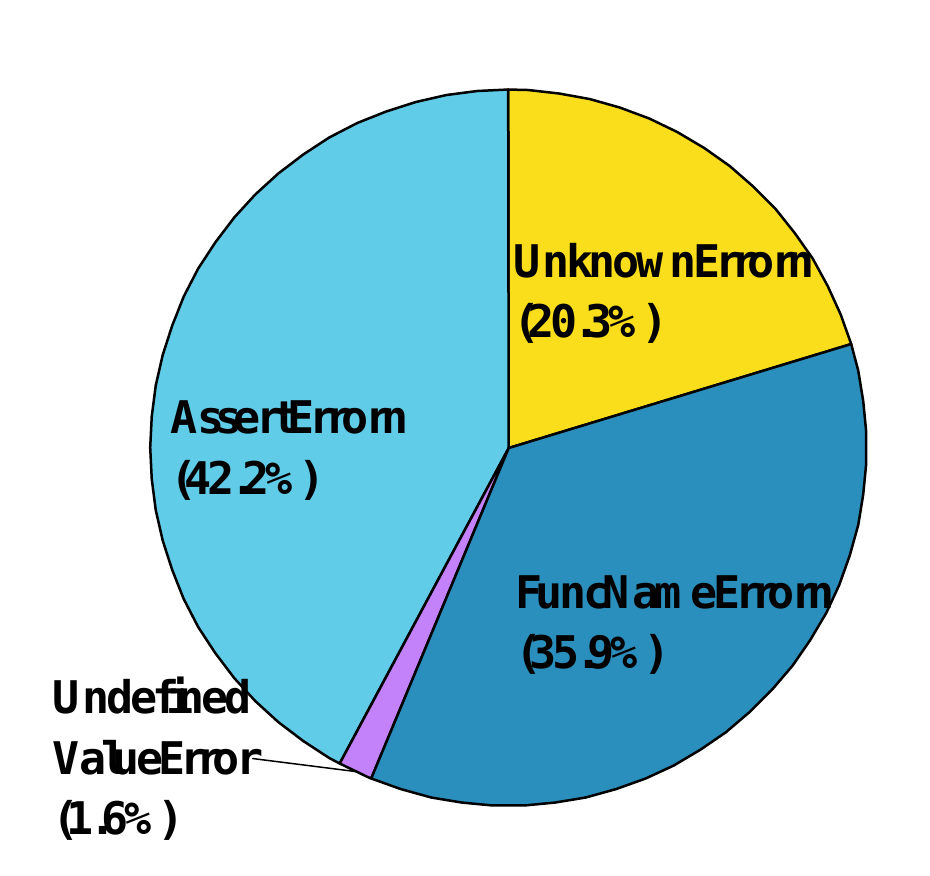}}
        \subfloat[][\SystemName with five-pass feedback refinement.]{\includegraphics[width=0.23\textwidth]{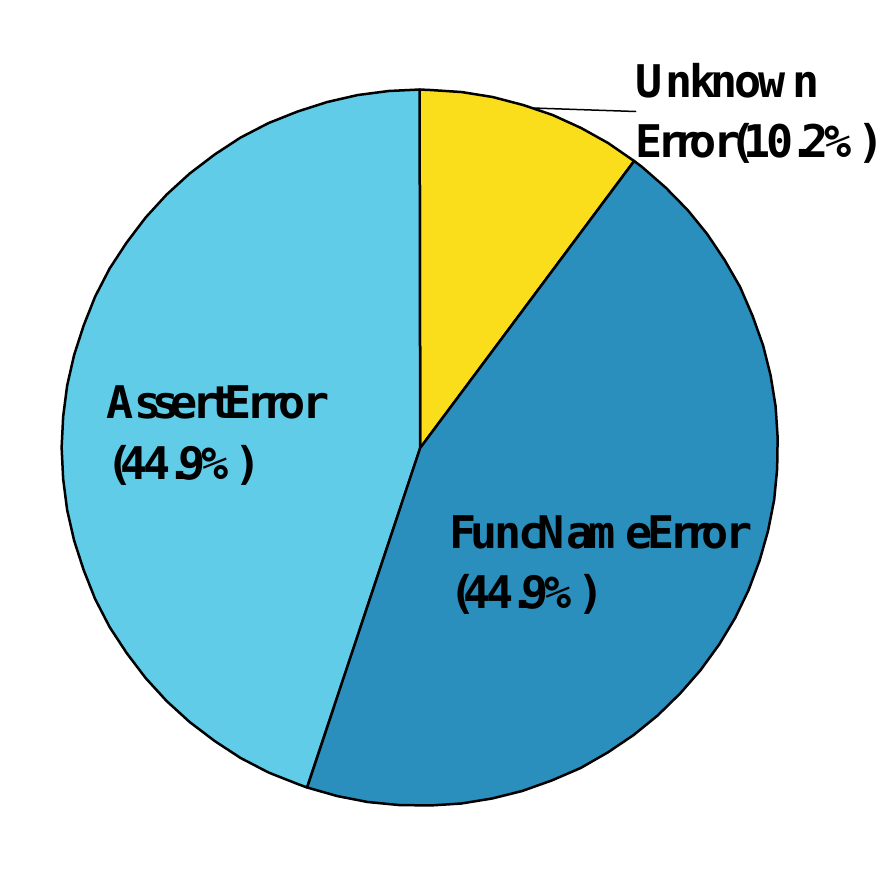}}
 	\caption{Distribution of error types in \SystemName under one-shot inference compared to one, three, and five iterations of feedback-directed refinement on the HumanEval-Decompile dataset (x86-64).}
 	 \vspace{-5mm}
 	\label{fig:error_analysis}
 \end{figure*}

\begin{table}[t]
\centering
\scriptsize
\caption{Performance on long functions with more than 200 instructions.}
\setlength{\tabcolsep}{6pt}
\begin{tabular}{lccc}
\toprule
\textbf{Method} & \textbf{Re-compilability (\%)} & \textbf{Re-executability (\%)} & \textbf{Edit-sim. (\%)} \\
\midrule
RetDec      & 91.1 & 10.7 & 23.9 \\
\SystemName & 100 & 76.9 & 44.4 \\
\bottomrule
\end{tabular}
\vspace{-1mm}

\label{tab:longfunc}
\end{table}

\begin{table}[t]
\centering
\scriptsize
    \caption{Performance under Different Test Case Sources on HumanEval-Decompile (x86-64). }
    \begin{tabular}{@{}c|ccc}
	  \toprule
		  \textbf{Test Case Setting}&\textbf{ Re-Comp (\%)}&\textbf{ Re-Exec (\%)}& \textbf{Edit-Sim (\%)}\\
        \midrule
         Structural Only (No Tests) &99.1 &88.0 &51.2  \\
         LLM-Generated Tests&\second{98.6} & \second{89.5}& \best{51.4}  \\
         Human-written Tests & \best{99.2} &\best{94.5} &\best{51.4}  \\
             \bottomrule
	\end{tabular}
 		\vspace{-2mm}
\label{tab:testcases-impact}

\end{table}

\subsection{Error Types Considered in \SystemName}

To evaluate the effectiveness of the \SystemName iterative refinement scheme, we categorize the errors encountered during the lifting process into four primary types:\textit{ FuncNameError}, \textit{AssertError}, \textit{Phi conflict}, and \textit{Definition Conflict }(Table~\ref{tab:ir_error_types} lists the details).
The Advisor--Validation loop progressively reduces structural failures and shifts the remaining failures toward behavioral mismatches. Early iterations prioritize compiler- and verifier-reported issues, which enables more candidates to pass IR validation and reach execution. For example, \textit{FuncNameError} decreases from 110 under one-shot inference to 22 by the third refinement pass, and \textit{PhiError} and \textit{DefinitionConflict} exhibit similar reductions. By the fifth iteration, the total failure count drops by 81.2\%, and the dominant remaining failures are \textit{AssertError}. This shift indicates that refinement moves the pipeline from failing at compilation/verification to reaching runtime checking, where the remaining discrepancies are primarily semantic.

\subsection{Impact of Test-Case–Guided Feedback}
\label{sec:testcase_impact}

Table~\ref{tab:testcases-impact} quantifies how functional test cases affect our iterative refinement. We compare two settings: (i) \emph{structural-only} refinement, where the loop relies on verifier/compilation checks and structural feedback without runtime tests, (ii) LLM-generated tests, and (iii) refinement with \emph{human-written} unit tests.

For the \emph{LLM-generated tests} setting, we follow an automated pipeline that first recovers a plausible calling interface for each target function and then synthesizes executable test drivers. In  detail, \texttt{angr} performs control-flow analysis, calling-convention inference, and candidate prototype generation for the target function, producing structured static facts. We then use DynamoRIO \texttt{drwrap}~\cite{bruening2004efficient} for function-level dynamic instrumentation, executing the function under a small set of heuristic inputs to collect parameter access patterns, return-type signals, and pointer/string behaviors, which form a dynamic observation report. Next, we package the disassembly, static facts, and dynamic observations into a structured prompt for DeepSeek-V3.2, which outputs multiple candidate signatures with confidence scores. These candidates are filtered and validated in two stages: we first discard signatures that contradict strong dynamic observations, and then run \texttt{libFuzzer}-based~\cite{serebryany2016continuous} feasibility checks by compiling each candidate into a harness that invokes the original binary and accepting it only if it remains crash-free under large volumes of randomized inputs. If all candidates fail, we feed the failure evidence back to the model for regeneration. Finally, for verified signatures, we perform differential checking by executing the original binary function and the lifted IR on the same inputs and comparing return values and observable memory side effects, and the discrepancies are recorded as runtime evidence to guide subsequent repairs.

We can observe that the re-compilability is high in all cases (98.6--99.2\%), indicating that toolchain checks and structural constraints are sufficient to produce verifier-clean IR regardless of the test source. In contrast, re-executability varies substantially with the availability and quality of tests. Using no tests yields 88.0\% re-executability, while adding LLM-generated tests improves it to 89.5\% (+1.5 points), and human-written tests further raise it to 94.5\% (+6.5 points over structural-only). This pattern suggests that tests primarily contribute behavioral constraints that resolve residual semantic ambiguity that is not visible to compilation and CFG-level checks. 
Edit similarity remains essentially unchanged (51.2--51.4\%), implying that the gains in re-executability are not driven by broader rewrites toward the reference text, but by localized corrections that improve behavior. Overall, the results indicate that \SystemName remains effective even without tests, but stronger functional oracles yield materially better semantic fidelity, while LLM-generated tests provide a lightweight proxy when developer-written tests are unavailable.

\subsection{Cost Analysis of \SystemName}
Table~\ref{tab:cost} reports the average token usage and cost of \SystemName instantiated with the Gemini-3 backbone on HumanEval (x86-64), broken down by inference sessions. The initial synthesis stage consumes 2{,}832 input tokens and 1{,}711 output tokens on average (\$0.0065). Each subsequent refinement step incurs a similar per-iteration cost, with inputs around 3.26K--3.42K tokens and outputs around 1.91K--2.06K tokens (\$0.0074--\$0.0079 per step). Over the full six-session run (initial draft plus five refinement rounds), the total averages 19{,}573 input tokens and 11{,}610 output tokens, corresponding to \$0.0446 per function. In practice, the full five-round refinement is primarily needed for harder cases, many functions converge in fewer iterations. Consequently, the average cost reported here should be interpreted as an upper-bound configuration under a fixed iteration budget rather than the typical cost for all inputs.

\begin{table}[t]
\centering
\scriptsize
\caption{Average token usage and monetary cost across inference sessions in \SystemName on HumanEval (x86-64).}
\setlength{\tabcolsep}{6pt}
\begin{tabular}{lrrr}
\toprule
\textbf{Inference session} & \textbf{Input tokens} & \textbf{Output tokens} & \textbf{Cost (\$)} \\
\midrule
Session 1 (Initial)   & 2832.49 & 1710.90 & 0.006548933 \\
Session 2 (IR 1)      & 3404.53 & 2060.09 & 0.007882541 \\
Session 3 (IR 2)      & 3418.04 & 2016.53 & 0.007758606 \\
Session 4 (IR 3)      & 3378.10 & 1987.15 & 0.007650497 \\
Session 5 (IR 4)      & 3281.11 & 1926.72 & 0.007420707 \\
Session 6 (Final)     & 3258.95 & 1908.56 & 0.007355145 \\
\midrule
\textbf{Total}        & \textbf{19573.21} & \textbf{11609.94} & \textbf{0.044616429} \\
\bottomrule
\end{tabular}
\vspace{-1mm}

\label{tab:cost}
\end{table}

\section{Discussion and Limitations}

We show that \SystemName generalizes across x86-64 and ARM64 in our evaluation. Extending to additional architectures (e.g., MIPS or RISC-V) requires rebuilding the structure-aware RAG database for the target ISA and toolchain configuration. While this adds engineering cost, the data construction pipeline is automated and can be rerun with the appropriate compiler backend and instrumentation settings. Nevertheless, retrieval quality and repair behavior may vary with ISA-specific idioms (e.g., branch forms, calling conventions, and jump-table lowering), so cross-ISA transfer is not guaranteed without reindexing and calibration.
Our evaluation on HumanEval-Decompile and MBPP benefits from curated unit tests, whereas real reverse-engineering scenarios (e.g., malware, proprietary libraries, or firmware) rarely provide input–output oracles. In such settings, functional feedback can be approximated by integrating binary test generation, for example via symbolic execution, concolic testing, or fuzzing to produce inputs that exercise diverse paths. These techniques can supply partial oracles (e.g., differential consistency checks, crash/non-crash signals, or metamorphic properties) that still guide refinement, although they may be incomplete and can introduce noise.
Finally, \SystemName is constrained by input length. For large functions or binaries with extensive inlining and unrolling, disassembly and IR context can exceed the backbone model’s context window. Our current design mitigates this by segmenting code into CFG-aligned regions and performing retrieval and repair at region granularity, but very large or highly entangled functions may still require additional decomposition (e.g., loop/SCC regions or dominator-subtree partitioning) and interface contracts to stitch regions back together.

\section{Related Work}

\subsection{LLVM Intermediate Representations Recovery}
\label{sec:llvm-ir-recovery}
Binary-to-LLVM IR decompilation is crucial for program analysis and reverse engineering tasks, including malware analysis and software migration. Tools like McSema~\cite{mcsema}, LLVM-MCTOLL~\cite{10.1145/3316482.3326354}, and RevGen~\cite{revgen} lift binaries to LLVM IR but struggle with Static Single Assignment (SSA) reconstruction, control flow preservation, and semantic fidelity under aggressive optimizations (e.g., -O3). For instance, McSema, while robust for x86 binaries, often produces IR with low structural similarity due to imprecise control flow recovery. Similarly, LLVM-MCTOLL generates functionally correct but structurally divergent IR, limiting its utility for high-precision analysis. Academic efforts, such as Schwartz et al.~\cite{10.5555/2534766.2534797}, propose control flow structuring but lack generalizability across optimization levels. LeanBin~\cite{wodiany2024leanbin} introduces mechanisms for binary lifting with a focus on debloating, addressing semantic gaps in IR transformation. However, its emphasis on code size reduction limits its applicability to precision-focused tasks. Recent work on lifting assembly to intermediate representations~\cite{armengol-estape2024forklift} provides foundational insights but lacks scalability for complex binaries. \SystemName overcomes these challenges using a control-flow-aware RAG database and pseudo-probe instrumentation, achieving high CFG similarity and edit similarity across diverse optimization levels.

\subsection{Learning-Based and Neural Decompilation Methods}
\label{sec:neural-decompilation}
Neural-based decompilation has gained traction with the use of transformers and large language models (LLMs). Methods like NeuralDecompile~\cite{10.1145/3428293}, LLM4Decompile~\cite{tan2024llm4decompiledecompilingbinarycode}, and SALDE~\cite{armengolestape2024sladeportablesmalllanguage} excel in binary-to-C decompilation, inferring high-level semantics. However, adapting LLMs for binary-to-LLVM IR lifting is challenging, as IR requires precise typing and structural fidelity to compiler formats, unlike human-readable C code. Meta’s LLMDecompiler FTD~\cite{cummins2024metalargelanguagemodel} and Forklift~\cite{armengol-estape2024forklift} target decompilation from Clang \texttt{-S} assembly to LLVM IR, leveraging metadata (e.g., symbol tables) unavailable in raw binaries, making it unsuitable for true binary decompilation. \SystemName integrates LLMs with a structure-aware RAG framework and iterative static-guided refinement, achieving robust IR reconstruction across optimization levels (-O0 to -O3). It outperforms baselines like McSema and LLM4Decompile in structural fidelity, particularly for complex binaries with system calls or inline assembly.

\subsection{Retrieval-Augmented Generation in Code Tasks}
\label{sec:rag-code}
Retrieval-Augmented Generation (RAG)~\cite{lewis2021retrievalaugmentedgenerationknowledgeintensivenlp} enhances generative models by incorporating external context, proving effective in code completion~\cite{parvez-etal-2021-retrieval-augmented} and program synthesis~\cite{10.5555/3327546.3327670}. However, its application to binary-to-LLVM IR decompilation remains underexplored. Existing RAG-based methods, such as CodeRAG~\cite{wang-etal-2025-coderag}, focus on high-level code synthesis and lack structure-aware retrieval for low-level IR. Recent advances like BTR~\cite{cao2024btrbinarytokenrepresentations} introduce efficient token-based retrieval for binary tasks, but they do not address IR-specific structural constraints. R2AG~\cite{ye2024r2agincorporatingretrievalinformation} and LI-RAGE~\cite{lin2024liragelateinteractionretrieval} propose advanced retrieval-integration architectures, offering potential for improving iterative retrieval-generation synergy. Similarly, DecLLM~\cite{wong2025decllm} and ITER-RETGEN~\cite{shao2023enhancing} emphasize iterative feedback loops, aligning with \SystemName’s approach. Specifically, DecLLM focuses on recompilable C code generation via LLMs and applies intermediate lifting to IR during training. \SystemName is fundamentally different in its target (LLVM IR vs. C) and setting (binary input vs. assembly from Clang -S), and architecture. \SystemName pioneers RAG for binary-to-LLVM IR lifting, using a fine-grained, control-flow-aware database to retrieve precise IR constructs, outperforming neural baselines in CFG and edit similarity, even for large binaries.

\section{Conclusion}
We have presented \SystemName, a multi-agent framework for lifting heavily optimized and stripped binaries into \emph{canonical, pre-optimized} LLVM IR. Rather than reconstructing source-level syntax, \SystemName targets verifier-clean, analysis-ready IR by combining control-flow-aware retrieval with verification-guided refinement. Across HumanEval-Decompile and MBPP, three ISAs (x86-64, ARM64, and RISC-V), and optimization levels from \texttt{-O0} to \texttt{-O3}, \SystemName consistently improves re-compilability, re-executability, and similarity to the reference IR over both rule-based and LLM-based baselines. Beyond matching the reference more closely, the lifted IR better supports downstream uses, including SVF-based program analysis and cross-ISA software migration. These results suggest that targeting compiler IR, rather than decompiled source code alone, provides a practical foundation for binary recovery that is both semantically useful and directly compatible with existing analysis and compilation toolchains.

    \bibliographystyle{IEEEtran}
    \bibliography{refs}

@article{tan2025sk2decompile,
  title={SK2Decompile: LLM-based Two-Phase Binary Decompilation from Skeleton to Skin},
  author={Tan, Hanzhuo and Li, Weihao and Tian, Xiaolong and Wang, Siyi and Liu, Jiaming and Li, Jing and Zhang, Yuqun},
  journal={arXiv preprint arXiv:2509.22114},
  year={2025}
}

@inproceedings{10.1145/3520312.3534867,
author = {Armengol-Estap\'{e}, Jordi and Woodruff, Jackson and Brauckmann, Alexander and Magalh\~{a}es, Jos\'{e} Wesley de Souza and O'Boyle, Michael F. P.},
title = {ExeBench: An ML-Scale Dataset of Executable C Functions},
year = {2022},
booktitle = {Proceedings of the 6th ACM SIGPLAN International Symposium on Machine Programming},
pages = {50–59},
}

@article{johnson2019billion,
  title={Billion-scale similarity search with {GPUs}},
  author={Johnson, Jeff and Douze, Matthijs and J{\'e}gou, Herv{\'e}},
  journal={IEEE Transactions on Big Data},
  volume={7},
  number={3},
  pages={535--547},
  year={2019}
}

@inproceedings{10.1145/3316482.3326354,
    author = {Yadavalli, S. Bharadwaj and Smith, Aaron},
    title = {Raising Binaries to LLVM IR with MCTOLL (WIP Paper)},
    year = {2019},
    booktitle = {Proceedings of the 20th ACM SIGPLAN/SIGBED International Conference on Languages, Compilers, and Tools for Embedded Systems},
    pages = {213–218}
 }

@article{wong2025decllm,
  title={DecLLM: LLM-Augmented Recompilable Decompilation for Enabling Programmatic Use of Decompiled Code},
  author={Wong, Wai Kin and Wu, Daoyuan and Wang, Huaijin and Li, Zongjie and Liu, Zhibo and Wang, Shuai and Tang, Qiyi and Nie, Sen and Wu, Shi},
  journal={Proceedings of the ACM on Software Engineering},
  volume={2},
  number={ISSTA},
  pages={1841--1864},
  year={2025},
  publisher={ACM New York, NY, USA}
}

@inproceedings{cummins2024metalargelanguagemodel,
author = {Cummins, Chris and Seeker, Volker and Grubisic, Dejan and Roziere, Baptiste and Gehring, Jonas and Synnaeve, Gabriel and Leather, Hugh},
title = {LLM Compiler: Foundation Language Models for Compiler Optimization},
year = {2025},
booktitle = {Proceedings of the 34th ACM SIGPLAN International Conference on Compiler Construction},
pages = {141–153},
series = {CC '25}
}

@misc{deepseek_chat,
  title        = {DeepSeek Chat},
  author       = {{DeepSeek}},
  howpublished = {\url{https://chat.deepseek.com/}},
  note         = {Accessed: 2025-01-24},
  year = {2025}
}

@misc{GoogleGemini,
  title = {{Gemini: Google's AI Model}}, 
  author = {{Google}}, 
  howpublished = {\url{https://gemini.google.com/}}, 
  note         = {Accessed: 2025-01-24},
  year = {2025}
}

@misc{chatgpt2025,
  author       = {{OpenAI}},
  title        = {{ChatGPT}},
  year         = {2025},
  howpublished = {\url{https://chatgpt.com/}},
  note         = {Accessed: 2025-01-24}
}

@misc{qwen_website,
  author       = {{Tongyi-Lab}},
  title        = {Qwen},
  howpublished = {\url{https://chat.qwen.ai/}},
  note         = {Accessed: 2025-01-24},
  year = {2025}
}

@misc{claude,
  title        = {Claude},
  author       = {{Anthropic}},
  year         = {2024},
  howpublished = {\url{https://www.anthropic.com/claude}},
  note         = {Accessed: 2025-01-24}
}

@inproceedings{lewis2021retrievalaugmentedgenerationknowledgeintensivenlp,
 author = {Lewis, Patrick and Perez, Ethan and Piktus, Aleksandra and Petroni, Fabio and Karpukhin, Vladimir and Goyal, Naman and K\"{u}ttler, Heinrich and Lewis, Mike and Yih, Wen-tau and Rockt\"{a}schel, Tim and Riedel, Sebastian and Kiela, Douwe},
 booktitle = {Advances in Neural Information Processing Systems (NeurIPS)},
 pages = {9459--9474},
 title = {Retrieval-Augmented Generation for Knowledge-Intensive NLP Tasks},
 volume = {33},
 year = {2020}
}

@inproceedings{parvez-etal-2021-retrieval-augmented,
    title = "Retrieval Augmented Code Generation and Summarization",
    author = "Parvez, Md Rizwan  and
      Ahmad, Wasi  and
      Chakraborty, Saikat  and
      Ray, Baishakhi  and
      Chang, Kai-Wei",
    booktitle = "Findings of the Association for Computational Linguistics: EMNLP 2021",
    year = "2021",
    pages = "2719--2734",
}

@inproceedings{10.5555/3327546.3327670,
author = {Hashimoto, Tatsunori B. and Guu, Kelvin and Oren, Yonatan and Liang, Percy},
title = {A retrieve-and-edit framework for predicting structured outputs},
year = {2018},
booktitle = {Proceedings of the 32nd International Conference on Neural Information Processing Systems (NeurIPS)},
pages = {10073–10083}
}

@inproceedings{armengolestape2024sladeportablesmalllanguage,
  author={Armengol-Estapé, Jordi and Woodruff, Jackson and Cummins, Chris and O'Boyle, Michael F.P.},
  booktitle={2024 IEEE/ACM International Symposium on Code Generation and Optimization (CGO)}, 
  title={SLaDe: A Portable Small Language Model Decompiler for Optimized Assembly}, 
  year={2024},
  pages={67-80}
}

@inproceedings{tan2024llm4decompiledecompilingbinarycode,
    title = "{LLM}4{D}ecompile: Decompiling Binary Code with Large Language Models",
    author = "Tan, Hanzhuo  and
      Luo, Qi  and
      Li, Jing  and
      Zhang, Yuqun",
    booktitle = "Proceedings of the 2024 Conference on Empirical Methods in Natural Language Processing (EMNLP)",
    year = "2024",
    pages = "3473--3487"
}

@article{austin2021program,
  title={Program synthesis with large language models},
  author={Austin, Jacob and Odena, Augustus and Nye, Maxwell and Bosma, Maarten and Michalewski, Henryk and Dohan, David and Jiang, Ellen and Cai, Carrie and Terry, Michael and Le, Quoc and others},
  journal={arXiv preprint arXiv:2108.07732},
  year={2021}
}

@inproceedings{shoshitaishvili2016state,
    author={Shoshitaishvili, Yan and Wang, Ruoyu and Salls, Christopher and Stephens, Nick and Polino, Mario and Dutcher, Andrew and Grosen, John and Feng, Siji and Hauser, Christophe and Kruegel, Christopher and Vigna, Giovanni},
  booktitle={2016 IEEE Symposium on Security and Privacy (SP)}, 
  title={SOK: (State of) The Art of War: Offensive Techniques in Binary Analysis}, 
  year={2016},
  pages={138-157}
}

@inproceedings{jiang2025novagenerativelanguagemodels,
  author       = {Nan Jiang and
                  Chengxiao Wang and
                  Kevin Liu and
                  Xiangzhe Xu and
                  Lin Tan and
                  Xiangyu Zhang and
                  Petr Babkin},
  title        = {Nova: Generative Language Models for Assembly Code with Hierarchical
                  Attention and Contrastive Learning},
  booktitle    = {The Thirteenth International Conference on Learning Representations (ICLR) 2025},
  year         = {2025}
}

@misc{mcsema,
  author       = {Lifting Bits},
  title        = {McSema: Lifting x86 and x86-64 binaries to LLVM bitcode},
  year         = {2021},
  howpublished = {\url{https://github.com/lifting-bits/mcsema}},
  note         = {Accessed: 2025-01-24}
}

@misc{revgen,
  author       = {{S2E Team}},
  title        = {Translating binaries to LLVM with RevGen},
  year         = {2020},
  howpublished = {\url{https://s2e.systems/docs/Tutorials/Revgen/Revgen.html}},
  note         = {Accessed: 2025-01-24}
}

@inproceedings{10.5555/2534766.2534797,
author = {Schwartz, Edward J. and Lee, JongHyup and Woo, Maverick and Brumley, David},
title = {Native \texttimes{}86 decompilation using semantics-preserving structural analysis and iterative control-flow structuring},
year = {2013},
booktitle = {Proceedings of the 22nd USENIX Conference on Security},
pages = {353–368}
}

@article{10.1145/3428293,
    title = {Neural Reverse Engineering of Stripped Binaries Using Augmented Control Flow Graphs},
    author = {David, Yaniv and Alon, Uri and Yahav, Eran},
    journal = {Proceedings of the ACM on Programming Languages},
    number = {OOPSLA},
    volume = {4},
    year = {2020}
}

@misc{llvmproject,
  author       = {{LLVM Project}},
  title        = {The LLVM Compiler Infrastructure Project},
  year         = {2025},
  howpublished = {\url{https://github.com/llvm/llvm-project}},
  note         = {Accessed: 2025-01-24.}
}

@inproceedings{armengol-estape2024forklift,
title={Forklift: An Extensible Neural Lifter},
author={Jordi Armengol-Estap{\'e} and Rodrigo C. O. Rocha and Jackson Woodruff and Pasquale Minervini and Michael O'Boyle},
booktitle={First Conference on Language Modeling (COLM)},
year={2024}
}

@misc{SciPyProceedings_11,
  author       = {Hagberg, Aric A. and Schult, Daniel A. and Swart, Pieter J.},
  title        = {NetworkX: Graph Isomorphism — NetworkX Documentation},
  year         = {2024},
  howpublished = {\url{https://networkx.org/documentation/stable/reference/algorithms/isomorphism.html}},
  note         = {Accessed: 2025-01-24}
}

@article{JMLR:v21:18-370,
  author  = {Giannis Siglidis and Giannis Nikolentzos and Stratis Limnios and Christos Giatsidis and Konstantinos Skianis and Michalis Vazirgiannis},
  title   = {GraKeL: A Graph Kernel Library in Python},
  journal = {Journal of Machine Learning Research},
  year    = {2020},
  volume  = {21},
  number  = {54},
  pages   = {1--5}
}

@inproceedings{wang-etal-2025-coderag,
    title = "{C}ode{RAG}-Bench: Can Retrieval Augment Code Generation?",
    author = "Wang, Zora Zhiruo  and
      Asai, Akari  and
      Yu, Xinyan Velocity  and
      Xu, Frank F.  and
      Xie, Yiqing  and
      Neubig, Graham  and
      Fried, Daniel",
    booktitle = "Findings of the Association for Computational Linguistics: NAACL 2025",
    year = "2025",
    pages = "3199--3214"
}

@inproceedings{wodiany2024leanbin,
	title        = {LeanBin: Harnessing Lifting and Recompilation to Debloat Binaries},
	author       = {Wodiany, Igor and Pop, Antoniu and Luj\'{a}n, Mikel},
	year         = 2024,
	booktitle    = {Proceedings of the 39th IEEE/ACM International Conference on Automated Software Engineering},
	pages        = {1434–1446}
}

@inproceedings{cao2024btrbinarytokenrepresentations,
 author = {Cao, Qingqing and Min, Sewon and Wang, Yizhong and Hajishirzi, Hannaneh },
 booktitle = {International Conference on Representation Learning},
 pages = {44229--44246},
 title = {BTR: Binary Token Representations for Efficient Retrieval Augmented Language Models},
 year = {2024}
}

@inproceedings{ye2024r2agincorporatingretrievalinformation,
    title = "{R}$^2${AG}: Incorporating Retrieval Information into Retrieval Augmented Generation",
    author = "Ye, Fuda  and
      Li, Shuangyin  and
      Zhang, Yongqi  and
      Chen, Lei",
    booktitle = "Findings of the Association for Computational Linguistics: EMNLP 2024",
    year = "2024",
    pages = "11584--11596"
}

@inproceedings{lin2024liragelateinteractionretrieval,
    title = "{LI}-{RAGE}: Late Interaction Retrieval Augmented Generation with Explicit Signals for Open-Domain Table Question Answering",
    author = "Lin, Weizhe  and
      Blloshmi, Rexhina  and
      Byrne, Bill  and
      de Gispert, Adria  and
      Iglesias, Gonzalo",
    booktitle = "Proceedings of the 61st Annual Meeting of the Association for Computational Linguistics (Volume 2: Short Papers)",
    year = "2023",
    pages = "1557--1566"
}

@inproceedings{shao2023enhancing,
title={Enhancing Retrieval-Augmented Large Language Models with Iterative Retrieval-Generation Synergy},
author={Zhihong Shao and Yeyun Gong and yelong shen and Minlie Huang and Nan Duan and Weizhu Chen},
booktitle={The 2023 Conference on Empirical Methods in Natural Language Processing},
year={2023},
pages = {9248--9274},
}

@INPROCEEDINGS{pseudo_probe,
  author={He, Wenlei and Yu, Hongtao and Wang, Lei and Oh, Taewook},
  booktitle={2024 IEEE/ACM International Symposium on Code Generation and Optimization (CGO)}, 
  title={Revamping Sampling-Based PGO with Context-Sensitivity and Pseudo-instrumentation}, 
  year={2024},
  pages={322-333},
}

@misc{retdec,
  author       = {Avast Software},
  title        = {RetDec: Open-source machine-code decompiler},
  year         = {2017},
  howpublished = {\url{https://github.com/avast/retdec}},
}

@inproceedings{
hu2022lora,
title={Lo{RA}: Low-Rank Adaptation of Large Language Models},
author={Edward J Hu and Yelong Shen and Phillip Wallis and Zeyuan Allen-Zhu and Yuanzhi Li and Shean Wang and Lu Wang and Weizhu Chen},
booktitle={International Conference on Learning Representations},
year={2022}
}

@article{VenkataKeerthy-2020-IR2Vec,
author = {VenkataKeerthy, S. and Aggarwal, Rohit and Jain, Shalini and Desarkar, Maunendra Sankar and Upadrasta, Ramakrishna and Srikant, Y. N.},
title = {{IR2Vec: LLVM IR Based Scalable Program Embeddings}},
year = {2020},
volume = {17},
number = {4},
journal = {ACM Trans. Archit. Code Optim.}
}

@article{bruening2004efficient,
  title={Efficient, transparent, and comprehensive runtime code manipulation},
  author={Bruening, Derek and Amarasinghe, Saman},
  year={2004},
  publisher={Massachusetts Institute of Technology, Department of Electrical Engineering~…}
}

@inproceedings{serebryany2016continuous,
  title={Continuous fuzzing with libfuzzer and addresssanitizer},
  author={Serebryany, Kosta},
  booktitle={2016 IEEE Cybersecurity Development (SecDev)},
  pages={157--157},
  year={2016},
  organization={IEEE}
}

@inproceedings{sui2016svf,
  title={SVF: interprocedural static value-flow analysis in LLVM},
  author={Sui, Yulei and Xue, Jingling},
  booktitle={Proceedings of the 25th international conference on compiler construction},
  pages={265--266},
  year={2016}
}

@inproceedings{zhu2026bridge,
  title = {Lifting Optimized Binaries to Canonical Compiler {IR} via Structure-Aware Retrieval and Iterative Verification},
  author = {Zhu, Xiaoao and Ren, Jie and Li, Zhiqiang and Zheng, Jie and Tang, Zhanyong and Wang, Zheng},
  booktitle = {Proceedings of the 64th Annual Meeting of the Association for Computational Linguistics (Volume 1: Long Papers)},
  year = {2026},
  month = jul,
  address = {San Diego, California, United States},
  publisher = {Association for Computational Linguistics},
  pages = {11498--11516},
  doi = {10.18653/v1/2026.acl-long.527},
  url = {https://aclanthology.org/2026.acl-long.527/}
}

\begin{IEEEbiographynophoto} {Xiaoao Zhu} received her BS degree in Computer Science from Shaanxi Normal University in 2023. Currently, she is pursuing the MS degree of software engineering at Shaanxi Normal University. Her research interest is mainly focused on LLVM optimization.
\end{IEEEbiographynophoto}

\begin{IEEEbiographynophoto} {Ni Zhang} received her BS degree in software engineering from Shaanxi Normal University in 2025. Currently, she is pursuing the MS degree of Computer Science  at Shaanxi Normal University. Her research interest is mainly focused on binary analysis.
\end{IEEEbiographynophoto}

\begin{IEEEbiographynophoto}	{Jie Ren} is an associate professor at the School of Computer Science at Shaanxi Normal University, Xi'an, China. He received the PhD degree in computer system architecture from Northwest University, Xi'an, China, in 2017. His research mainly focuses on mobile system optimisation and runtime scheduling.
\end{IEEEbiographynophoto}

\begin{IEEEbiographynophoto}{Zhiqiang Li} received the Ph.D degree in Computer Application  Technology from Wuhan
University, Wuhan, China, in 2018. He is currently
an Associate Professor with the School of Computer Science,
Shaanxi Normal University, Xi’an, China. His main
research interests include intelligence software
engineering and machine learning.
\end{IEEEbiographynophoto}

\begin{IEEEbiographynophoto}{Jie Zheng} received the BS.
degree in communications engineering from Nanchang University in 2008 and the Ph.D. degree from
the Department of Telecommunications Engineering,
Xidian University, China, in 2014. He is currently
an Associate Professor with the Information Science
and Technology Institute, Northwest University,
Xi’an, China. His research interests include energy-efficient transmission, wireless resource allocation,
and edge intelligence.
\end{IEEEbiographynophoto}

\begin{IEEEbiographynophoto}{Zhanyong Tang} is currently a Professor with the
School of Information Science and Technology,
Northwest University. His current research interests include network and system security.
\end{IEEEbiographynophoto}

\begin{IEEEbiographynophoto}	{Zheng Wang}
is chair professor of Intelligent Software Technology at the School of Computing at the University of Leeds and a Turing Fellow at the Alan Turing Institute. His research focuses on accelerating large-scale deep learning models, program optimisation, applied machine learning and system security.
\end{IEEEbiographynophoto}

\end{document}